\documentclass[pdflatex,sn-mathphys-num,iicol]{sn-jnl}

\usepackage{graphicx}
\usepackage{amsmath,amssymb,amsfonts}
\usepackage{bm}
\usepackage[title]{appendix}
\usepackage{booktabs}

\newcommand{\Top}{\mathcal{T}}
\newcommand{\Mop}{\mathcal{M}}
\newcommand{\Pop}{\mathcal{P}}
\newcommand{\Xc}{\mathcal{X}}
\newcommand{\EFc}{E_{\mathrm F}^{c}}
\newcommand{\RH}{R_{\mathrm H}}
\newcommand{\RL}{R_{\mathrm L}}
\newcommand{\VH}{V_{\mathrm H}}
\newcommand{\Vconf}{V_{\mathrm{conf}}}
\newcommand{\Udis}{U_{\mathrm{dis}}}
\newcommand{\nel}{n_{\mathrm{el}}}
\newcommand{\scblin}{\textsc{scblin}}

\begin{document}

\title[The quantized Hall effect knows which way the current flows]{The quantized Hall effect knows which way the current flows: mirror covariance and rectification in a laterally asymmetric Hall bar}

\author*[1]{\fnm{Afif} \sur{Siddiki}}\email{afif.siddiki@atlas.edu.tr}

\affil*[1]{\orgdiv{Vocational School}, \orgname{Atlas University}, \orgaddress{\city{\.{I}stanbul}, \postcode{34408}, \country{T\"urkiye}}}

\abstract{The quantized Hall effect is the textbook emblem of universality: a resistance fixed by $h/e^2$, apparently indifferent to the shape of the sample, to its edges and to the direction of the current. That indifference is only half true. The plateau value is universal; the plateau is not. In a gate-defined Hall bar with one hard and one soft edge, reversing the current moves the plateau edge, and exchanging the edges reverses the move. We identify the symmetry that forces this. Because a perpendicular field breaks time reversal, the relevant operation is the in-plane $C_2$ rotation; within the self-consistent screening model it gives the covariance relation $R_\alpha[B,V_{\mathrm{conf}}(x),I]=R_\alpha[B,V_{\mathrm{conf}}(-x),-I]$, exact deep in the nonlinear regime and independent of linear response or microscopic reversibility. It predicts a rectification that is odd under exchange of the gate voltages and reverses with the magnetic field. The field-reversal test leaves the frozen lateral disorder untouched, so it holds for a single sample; it has not yet been performed. A further measurement, collapsing one incompressible strip, discriminates between co-directed and chiral pictures of the current. Self-consistent calculations show the soft-edge strip widening from 69 to 106\,nm on current reversal, moving the Hall resistance from 0.940 to $0.999994\,h/2e^2$ within a bounded window in magnetic field. The effect is a gate-tunable vector-product nonreciprocity, not magnetochiral anisotropy.}


\maketitle

\section{Introduction}

Few results in condensed-matter physics carry the authority of the integer quantized Hall effect (IQHE). Its resistance, $h/Ne^2$, is reproduced across materials, laboratories and decades with a precision that now realises the ohm. That precision has fostered a stronger and rarely examined belief: that the effect is indifferent to the particulars of the sample --- to the shape of its edges, to the electrostatics of its confinement, and to the direction in which the current is driven. In the Landauer--B\"uttiker (LB) description~\cite{Buttiker1988} that indifference is built in from the start. The current is carried by one-dimensional edge channels whose propagation direction is locked to $\operatorname{sgn}B$, the response follows from counting transmitted channels, and $\RH=h/Ne^2$, $\RL=0$ irrespective of the lateral confinement and of the sign of the imposed current.

This paper argues that the belief is half right, and that the half which is wrong is within experimental reach. The plateau \emph{value} is universal; the plateau \emph{itself} is not. Where quantization sets in and where it is lost are decided by the boundary conditions that the gates impose and by the sign of the current. A Hall bar with one hard and one soft edge is not the same conductor for $+I$ and for $-I$: in the calculation of Sec.~\ref{sec:mech} it is quantized to six digits in one direction and falls six per cent short in the other, at the same magnetic field and the same magnitude of current (Table~\ref{tab:resp}). The quantized Hall effect, in other words, knows which way the current flows, and it knows what the edges look like.

The screening theory starts elsewhere~\cite{Siddiki2004PRB,Siddiki2004IJMPB,Guven2003,Gerhardts2008}. The direct Coulomb interaction, the actual confinement and the imposed non-equilibrium current are treated self-consistently; the electron system fragments into compressible and incompressible regions, and the dissipation-suppressed current is carried by the incompressible strips (ISs). That fragmentation is not hypothetical: it has been mapped by scanning-force measurements of the Hall potential~\cite{Ahlswede2001} and by local compressibility imaging~\cite{Ilani2004}, and the predicted migration of the current path from the edges into the bulk within a single plateau has recently been detected with an ohmic probe at the centre of a Hall bar~\cite{Sirt2025APL}.

One experimental observation motivates this paper. In a gate-defined bar with one hard and one soft edge --- the soft edge being the one with the longer depletion length and hence the shallower density gradient --- $\RL$ and $\RH$ measured at $+I$ and at $-I$ differ at finite current, and the roles of the two polarities are interchanged when the hard and soft edges are exchanged~\cite{Siddiki2009EPL,Siddiki2010NJP}. This is a rectification: a response that is not even in the current. Our purpose is to show that it is a nonlinear transport diagnostic of broken lateral inversion symmetry --- broken $\Pop$ is necessary but not sufficient, the effect requiring in addition a finite current and the current-induced electrostatic redistribution that carries it --- to derive the relation that makes this precise, and to extract from it a field-reversal test that has not yet been performed.

What is new here needs stating explicitly, because neither the rectification nor the screening calculation that reproduces it is new. Both belong to a body of work published between 2004 and 2010~\cite{Siddiki2004PRB,Siddiki2004IJMPB,Siddiki2007IJMPB,Siddiki2008PhysE,Siddiki2009EPLb,Siddiki2009EPL,Siddiki2010NJP}, in which the self-consistent screening description of the incompressible strips was established, the range dependence of disorder was examined for the plateau widths, and the polarity-dependent asymmetry of a hard--soft bar was both measured and computed. The existence of a current-direction asymmetry in a hard--soft bar is not a result of this paper. Three things are. First, the effect is shown to be forced by a single broken spatial operation rather than reproduced by a fitted calculation: Eq.~\eqref{eq:cov} below is a covariance of the nonlinear self-consistent equations, following from the evenness of the Poisson kernel and a relabelling of coordinates, so that it constrains the response without linearisation and without microscopic reversibility. Second, that covariance yields a field-reversal relation, Eq.~\eqref{eq:Bodd}, which has not been tested, and which we show to be better conditioned against frozen disorder than the gate-exchange relation already examined in Ref.~\cite{Siddiki2010NJP}. Third, the effect is placed within the classification of nonreciprocal responses, Eq.~\eqref{eq:class}, and separated by explicit criteria from magnetochiral anisotropy, from Berry-curvature-dipole nonlinearity, and from the gate- and contact-driven nonreciprocity of Ref.~\cite{Iwakiri2023}. The numerics of Sec.~\ref{sec:mech} serve these three points; they are not offered as a new calculation of the effect itself.

What they are offered as evidence for is correspondingly narrow, and we state it here rather than only in Sec.~\ref{sec:conv}: the sign of the effect, its behaviour under the operations of Sec.~\ref{sec:ops}, and the existence of a bounded window in field. The magnitude at a given field, and the position of that window, depend on the averaging length and on the disorder model, and are not put forward as predictions.

\section{Model and mirror covariance}

\subsection{The quasi-local screening model}\label{sec:model}

For a translationally invariant bar of width $2d$ along $y$ with $\bm B=B\hat{\bm z}$,
\begin{equation}
\begin{split}
H={}&\sum_i\left[\frac{\bigl(\bm p_i+e\bm A(\bm r_i)\bigr)^2}{2m^*}+\Vconf(x_i)+V_{\mathrm H}(x_i)\right]\\
&+H_{\mathrm{dis}}+H_{\mathrm Z},
\end{split}
\label{eq:H}
\end{equation}
with $\nabla\times\bm A=\bm B$. The Zeeman term $H_{\mathrm Z}=\sigma g^*\mu_{\mathrm B}B$ is covariant under all three operations of Sec.~\ref{sec:ops}: both the spin and $\bm B$ are axial, so $\Top$ and $\Mop$ reverse each of them and $H_{\mathrm Z}$ is invariant, while $\Pop$ leaves both unchanged. Including spin therefore does not modify Eqs.~\eqref{eq:cov}--\eqref{eq:Bodd}. The numerical work below sets $g^*=0$ and treats the levels as spin degenerate, so the computed filling factors are even; the field-reversal derivation is unaffected, but the quantitative results do not apply to odd $\nu$, where exchange enhancement controls the gap. Electrostatic self-consistency is imposed through
\begin{align}
V(x)&=\Vconf(x)\nonumber\\
&\quad+\frac{2e^2}{\bar\kappa}\int_{-d}^{d}dx'\,K(x,x')\bigl[\nel(x')-n_0\bigr],\label{eq:V}\\
K(x,x')&=\ln\left|\frac{\sqrt{(d^2-x^2)(d^2-x'^2)}+d^2-xx'}{(x-x')d}\right|,\label{eq:K}\\
\nel(x)&=\int dE\,D(E)\,f\bigl(E+V(x)-\mu^*(x)\bigr),\label{eq:n}
\end{align}
with $n_0$ the fixed donor background, $D(E)$ the collision-broadened Landau density of states and $\mu^*(x)$ the electrochemical potential; $\nu(x)=2\pi\ell_B^2\nel(x)$. Transport follows from a quasi-local Ohm's law,
\begin{equation}
\nabla\mu^*(x)/e=\hat\rho(x)\,\bm j(x),\quad j_x\equiv0,\quad E_y(x)\equiv E^0_y,
\label{eq:ohm}
\end{equation}
so that
\begin{equation}
j_y(x)=\frac{E^0_y}{\rho_l(x)},\qquad E_x(x)=E^0_y\frac{\rho_{\mathrm H}(x)}{\rho_l(x)},
\label{eq:j}
\end{equation}
\begin{equation}
\begin{gathered}
I=E^0_y\int_{-d}^{d}\frac{dx}{\rho_l(x)},\qquad \VH=E^0_y\int_{-d}^{d}dx\,\frac{\rho_{\mathrm H}(x)}{\rho_l(x)},\\
\RH=\frac{\VH}{I},\qquad \RL=\frac{2dE^0_y}{I}.
\end{gathered}
\label{eq:R}
\end{equation}
The conductivity tensor entering Eq.~\eqref{eq:R} is averaged over a length $\lambda\sim\lambda_{\mathrm F}/2$ before inversion, which simulates the finite extent of the wave functions and is what gives the plateaus a finite width~\cite{Siddiki2004PRB}. Note that $\RL$ as defined in Eq.~\eqref{eq:R} is the drop over a length $2d$ along the bar, i.e.\ the resistance of one square; comparison with a four-terminal measurement carries the factor $L_{\mathrm{probe}}/2d$.

Two remarks on scope. First, Eq.~\eqref{eq:j} gives an imposed current density of one sign across the bar, since $E^0_y$ is a constant and $\rho_l>0$. We state this as a property of the prescription and not of the microscopic system: within the quasi-local model used here, the imposed transport-current component is of one sign across the bar. It does not follow that the microscopic system has no chiral edge structure. Sirt and Ludwig~\cite{SirtLudwig2026} decompose the total flow into imposed and closed-loop persistent components and call the flow chiral, a reading supported by multiterminal measurements~\cite{Sirt2025NJP}; a quantum-electrostatic calculation without the Thomas--Fermi approximation places part of the imposed current in the compressible regions~\cite{Armagnat2020}. The relation derived next does not require taking a position on that decomposition, since the persistent component follows from the same self-consistent problem, Eqs.~\eqref{eq:V}--\eqref{eq:n}, and transforms under a relabelling of coordinates exactly as the imposed component does. We none the less retain the co-directed transport current here, for reasons set out in Sec.~\ref{sec:LB}, where we also propose a measurement that would distinguish the two descriptions.

Second, a non-zero $\sigma_{\mathrm H}$ is odd under time reversal and odd under an in-plane mirror, and even under their product. The existence of the Hall effect is therefore already the statement that $\Top$ and $\Mop$ are individually broken while $\Mop\Top$ is not. What follows is the question of what breaks that composite.

\subsection{The operations}\label{sec:ops}

A symmetry operation must map the complete configuration $\Xc=\{B,\Vconf(x),n_0,I\}$ onto itself; transforming one term of Eq.~\eqref{eq:H} is not enough. Three generators suffice for the main argument (charge conjugation is deferred to Appendix~\ref{app:C}, for reasons explained there).

\paragraph{Time reversal $\Top$ (antiunitary)} $B\to-B$, $\bm j\to-\bm j$ hence $I\to-I$, $\Vconf$ unchanged.

\paragraph{Mirror parity $\Mop$ (unitary)} In two space dimensions the improper spatial operation with $\det=-1$ is a single reflection, $\Mop:(x,y)\to(-x,y)$. Since $\bm B$ is axial, an in-plane mirror reverses $B_z$: $B\to-B$, $\Vconf(x)\to\Vconf(-x)$, $I\to+I$, $E_x(x)\to-E_x(-x)$.

\begin{figure*}[t]
\centering
\includegraphics[width=\textwidth]{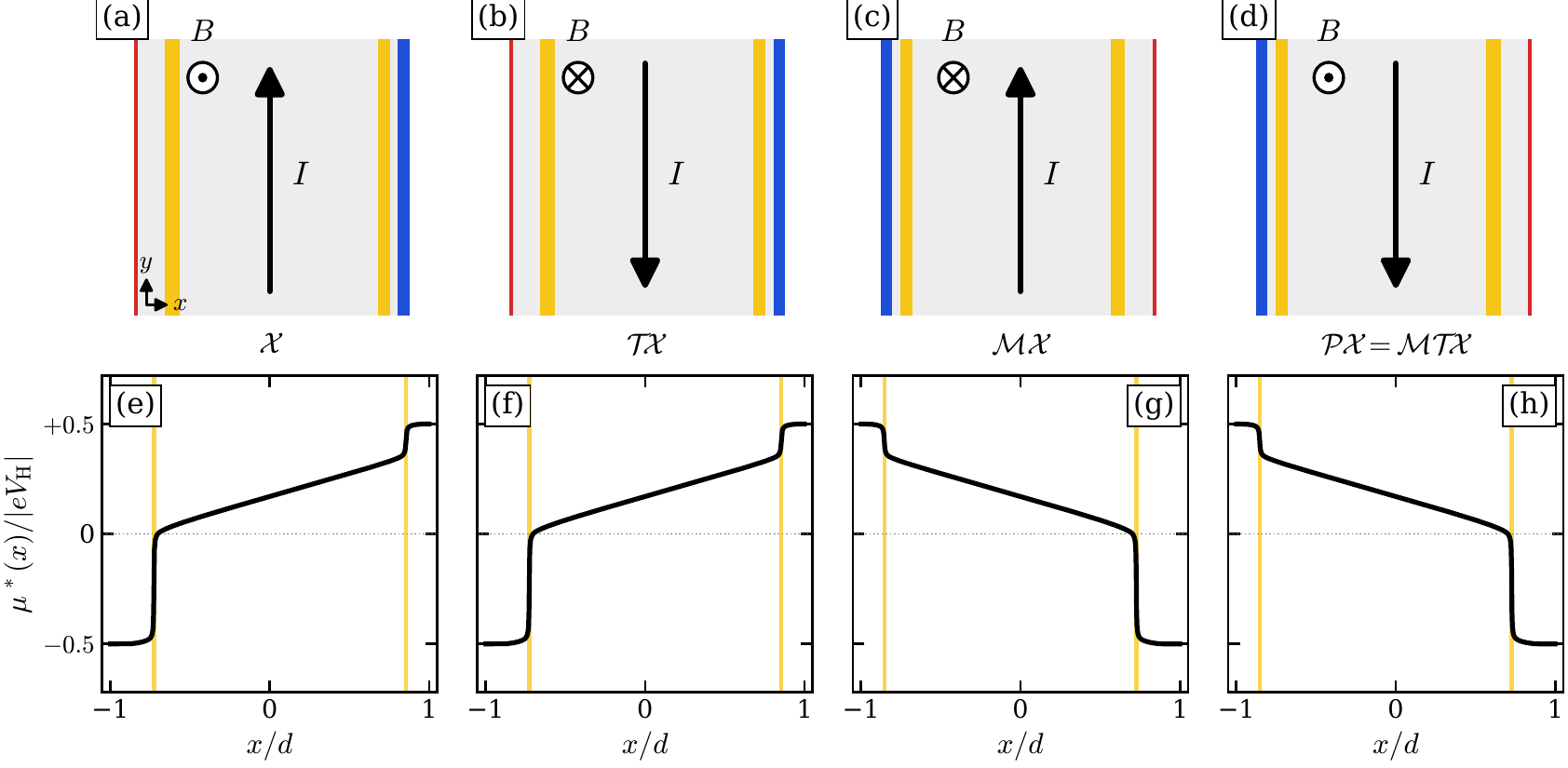}
\caption{Action of the discrete operations on a hard--soft Hall bar. Panels (a)--(d) are schematic; panels (e)--(h) are \emph{computed}, and are the self-consistent electrochemical potentials returned by the \scblin\ calculation of Sec.~\ref{sec:mech}, not drawings. (a)--(d), the configuration. The thin (red) edge is the soft, the thick (blue) edge the hard confinement, in the same arrangement as the calculation; shaded (yellow) regions are the $\nu=2$ incompressible strips, wide at the soft and narrow at the hard edge. (a) reference configuration $\Xc$; (b) $\Top\Xc$ reverses $B$ and $I$; (c) $\Mop\Xc$ reverses $B$ and interchanges the edges; (d) $\Pop\Xc=\Mop\Top\Xc$ keeps $B$, interchanges the edges and reverses the current. Of the three transformed configurations only (d) can be realised without reversing the magnet. The strips are drawn at four times their computed width so that the soft/hard contrast is visible; their positions are to scale. (e)--(h), the calculated electrochemical potential $\mu^*(x)$ of the configuration directly above, with the strips shaded at their computed widths. All four panels are divided by the same fixed positive quantity, the $|e\VH|$ of the reference run; no panel is normalised to its own signed Hall voltage. With that convention the transformation rules carry no minus signs. $\Top$ reverses $B$ and $I$ together, so $\rho_{\mathrm H}j_y$ and hence the whole Hall drop are unchanged and $\tilde\mu^*(x)=+\mu^*(x)$: panel (f) duplicates (e), which is precisely why $\Top$ on its own is of no use here. $\Mop$ is a reflection, under which the pseudovector $B_z$ reverses of itself, so the mirrored system carries the reflected profile, $\tilde\mu^*(x)=+\mu^*(-x)$; and likewise $\Pop$: $\tilde\mu^*(x)=+\mu^*(-x)$. Panel (e) is the converged solution of Eqs.~\eqref{eq:V}--\eqref{eq:R} for the reference device (depletion $300|150$\,nm, $e\VH=+0.2\EFc$, $\nu(0)=2.5$, $kT=5\times10^{-3}\EFc$, $N=800$, $\lambda=40$\,nm) and panel (h) a second, independent solution for the mirror-conjugate device ($150|300$\,nm) at reversed current; (f) and (g) follow from these by the rules above without further computation. That the computed (h) is the reflection of the computed (e) to $4\times10^{-5}$ of $|e\VH|$ --- the opposite sign being wrong by a full unit --- is Eq.~\eqref{eq:cov} read off two separate runs of the code. The residual tilt across the compressible bulk in (e) is physical rather than numerical: at this current the soft-edge strip has narrowed to 69\,nm and no longer carries the whole current, which is the $+I$ row of Table~\ref{tab:resp}.}
\label{fig:ops}
\end{figure*}

\paragraph{In-plane inversion $\Pop=\Mop\Mop_y$ (unitary)} $(x,y)\to(-x,-y)$ has $\det=+1$ in two dimensions --- it is a $C_2$ rotation about $\hat{\bm z}$ --- and therefore preserves $B_z$:
\begin{equation}
\Pop:\ \bigl(B,\Vconf(x),I\bigr)\longrightarrow\bigl(B,\Vconf(-x),-I\bigr).
\label{eq:P}
\end{equation}
The actions are collected in Table~\ref{tab:ops} and shown in Fig.~\ref{fig:ops}, whose lower row gives the calculated electrochemical potential of each transformed configuration. Note the coincidence
\begin{equation}
\Mop\Top:(B,\Vconf(x),I)\to(B,\Vconf(-x),-I)\equiv\Pop.
\label{eq:MT}
\end{equation}

\begin{table}[t]
\centering
\caption{Action of the spatial and time-reversal operations on the configuration $\Xc$. Charge conjugation is treated separately in Appendix~\ref{app:C}.}
\label{tab:ops}
\begin{tabular}{lccc}
\toprule
operation & $B$ & $\Vconf$ & $I$\\
\midrule
$\Top$ & $-B$ & $\Vconf(x)$ & $-I$\\
$\Mop$ & $-B$ & $\Vconf(-x)$ & $+I$\\
$\Pop=\Mop\Mop_y$ & $+B$ & $\Vconf(-x)$ & $-I$\\
$\Mop\Top\ (=\Pop)$ & $+B$ & $\Vconf(-x)$ & $-I$\\
\bottomrule
\end{tabular}
\end{table}

The antiunitary ``$\Pop\Top$'' operation and the unitary in-plane inversion act identically on the configuration. They are not interchangeable as theorems: $\Pop$ is a relabelling of coordinates and holds for any solution of Eqs.~\eqref{eq:V}--\eqref{eq:R}, whereas $\Mop\Top$ would additionally require microscopic reversibility, which a dissipative steady state does not possess. Everything below is derived from $\Pop$.

\subsection{The covariance relation}

The kernel~\eqref{eq:K} satisfies
\begin{equation}
K(x,x')=K(-x,-x'),
\label{eq:Keven}
\end{equation}
by inspection: both $\sqrt{(d^2-x^2)(d^2-x'^2)}+d^2-xx'$ and $|(x-x')d|$ are invariant under $(x,x')\to(-x,-x')$. Let $\{\nel,V,\mu^*,E^0_y\}$ solve Eqs.~\eqref{eq:V}--\eqref{eq:R} for confinement $\Vconf(x)$ and current $I$. By~\eqref{eq:Keven} the mirrored fields $\tilde n(x)=\nel(-x)$, $\tilde V(x)=V(-x)$, $\tilde\mu^*(x)=\mu^*(-x)$, $\tilde E^0_y=-E^0_y$ solve the same equations with $\Vconf(x)\to\Vconf(-x)$, and by~\eqref{eq:j} $\tilde j_y(x)=-j_y(-x)$, $\tilde E_x(x)=-E_x(-x)$. Hence $\tilde I=-I$, $\tilde V_{\mathrm H}=-\VH$, $\tilde E^0_y/\tilde I=E^0_y/I$, and
\begin{equation}
R_\alpha\bigl[B,\Vconf(x),I\bigr]=R_\alpha\bigl[B,\Vconf(-x),-I\bigr],
\label{eq:cov}
\end{equation}
with $\alpha=\mathrm H,\mathrm L$.
Equation~\eqref{eq:cov} is exact within the self-consistent quasi-local screening model under mirror-related boundary conditions. It uses no linearisation and no time reversal, so it survives the nonlinear feedback of the current on $\mu^*$ and $\nel$ and the presence of dissipation.

\paragraph{Scope of the covariance relations} Equations~\eqref{eq:cov}, \eqref{eq:M} and \eqref{eq:comp} hold for the complete transformed configuration $\{B,\Vconf(x),\Udis(x),\text{contacts},I\}\to\{B,\Vconf(-x),\Udis(-x),\text{mirrored contacts},-I\}$, not merely for $\Vconf(x)\to\Vconf(-x)$. Each is exact for the disorder-free or disorder-averaged quasi-local model and between two devices that are genuine mirror conjugates. In one physical sample, exchanging the gate voltages does not mirror the frozen impurities, the donors or the contacts, so Eqs.~\eqref{eq:cov} and \eqref{eq:M} hold up to sample-specific corrections and the experimental statement is that the antisymmetric component dominates, not that the symmetric component vanishes. Their composition, Eq.~\eqref{eq:comp}, is a special case and is treated separately below.

The disorder term of Eq.~\eqref{eq:H} requires a word, because the derivation transforms $\Vconf$ but not $H_{\mathrm{dis}}$. The relation as written presumes that the disorder is itself mirror covariant, and this is satisfied in three circumstances and in no others: (i) when disorder enters only as a spatially uniform level broadening, which is the case in the SCBA treatment used here, where the impurity range $R$ and strength $\gamma_I$ produce a $\Gamma$ that is the same at every $x$; (ii) after a disorder average, since the ensemble is statistically mirror symmetric even when no member of it is; or (iii) between two devices that are genuine mirror conjugates, $\Udis(x)$ and $\Udis(-x)$. For a single physical sample none of this is automatic: exchanging the two gate voltages mirrors the confinement but leaves the frozen impurity configuration, the donor positions and the local density fluctuations exactly where they were. Equation~\eqref{eq:cov} therefore holds for the idealised model, and for a real device it holds up to sample-specific disorder corrections whose size is set by the disorder-induced spread of the plateau edges. We return to the practical consequence in Sec.~\ref{sec:artefacts}.

Nor is Eq.~\eqref{eq:cov} an experimentally exact statement for other reasons: it presumes translational invariance along the bar, a device whose gates and contacts map onto each other under $(x,y)\to(-x,-y)$, and the quasi-local conductivity prescription itself. Contact asymmetry, nonlocal response, heating, breakdown dynamics and finite-length effects all lie outside it, and the range of validity is discussed in Sec.~\ref{sec:validity}.

The mirror $\Mop$ gives a second relation. Landau quantisation depends on $|B|$ only, so $\rho_l$ is unchanged under $B\to-B$ while $\rho_{\mathrm H}$ changes sign. This is exact for the Hamiltonian~\eqref{eq:H} and approximate for a real GaAs well, where the Dresselhaus and Rashba terms are omitted from Eq.~\eqref{eq:H}: they contribute a spin splitting of order 0.2--0.4\,meV at $k_{\mathrm F}=0.137$\,nm$^{-1}$, against $\hbar\omega_c=10.3$\,meV at 5.98\,T. We quote the ratio $\lesssim4\times10^{-2}$ as an upper bound on the available non-orbital energy scale rather than as a computed asymmetry: spin--orbit coupling does not by itself make $\rho_l$ odd in $B$, which requires it together with structural inversion breaking, so the number bounds what such a term could contribute without estimating it. Taking $\tilde n(x)=\nel(-x)$ and $\tilde E^0_y=+E^0_y$ gives $\tilde j_y(x)=j_y(-x)$, $\tilde E_x(x)=-E_x(-x)$, hence $\tilde I=I$, $\tilde V_{\mathrm H}=-\VH$ and
\begin{equation}
\begin{aligned}
\RL\bigl[B,\Vconf(x),I\bigr]&=\RL\bigl[-B,\Vconf(-x),I\bigr],\\
\RH\bigl[B,\Vconf(x),I\bigr]&=-\RH\bigl[-B,\Vconf(-x),I\bigr].
\end{aligned}
\label{eq:M}
\end{equation}
The same qualification applies: Eq.~\eqref{eq:M} transforms the Hall tensor and the spatial profile, and it equally requires the disorder, the boundaries and the contact labelling to transform. Composing~\eqref{eq:M} with~\eqref{eq:cov} eliminates the mirrored device:
\begin{equation}
\begin{split}\RL(B,+I)&=\RL(-B,-I),\\ \RH(B,+I)&=-\RH(-B,-I).\end{split}
\label{eq:comp}
\end{equation}
Equation~\eqref{eq:comp} is better conditioned against frozen disorder than either of the relations it is built from, and the reason is easily overlooked. Each factor moves the impurity potential: $\Pop$ takes $\Udis(x,y)\to\Udis(-x,-y)$ and $\Mop$ takes it to $\Udis(-x,y)$, so neither is realisable in a given sample. Composed, however, they return the lateral coordinate to itself,
\begin{equation}
\Udis(x,y)\xrightarrow{\ \Pop\ }\Udis(-x,-y)\xrightarrow{\ \Mop\ }\Udis(x,-y),
\label{eq:Udis}
\end{equation}
leaving only $y\to-y$, an exchange of source and drain. The model used here is translationally invariant along the bar, so $\Udis$ carries no $y$ dependence at all and Eq.~\eqref{eq:comp} is exact for a \emph{single frozen realisation}, not merely for the disorder-averaged ensemble. What remains to be assumed is therefore a condition on the contacts --- that the current and voltage probes map onto each other under source--drain exchange --- together with whatever genuine $y$ dependence the disorder of a real device possesses. It is not a condition on the lateral impurity configuration, which is precisely the quantity the gate-exchange relation~\eqref{eq:gex} cannot control. In this specific sense the field-reversal test is the cleaner of the two, and we return to the consequence in Sec.~\ref{sec:artefacts}.

\section{Experimental consequences}

\subsection{Exchange of the gate voltages}

Define the rectification
\begin{equation}
\Delta R_\alpha\equiv R_\alpha\bigl[\Vconf,+I\bigr]-R_\alpha\bigl[\Vconf,-I\bigr],
\label{eq:DR}
\end{equation}
which by Eq.~\eqref{eq:cov} equals
\begin{equation}
\Delta R_\alpha=R_\alpha\bigl[\Vconf(x),+I\bigr]-R_\alpha\bigl[\Vconf(-x),+I\bigr].
\label{eq:DR2}
\end{equation}
$\Delta R_\alpha$ is thus identically the difference of one observable evaluated on two mirror-conjugate devices. It vanishes for a completely mirror-symmetric configuration --- confinement, disorder and contacts alike, or the disorder-averaged model --- whatever the current and however nonlinear the response, which is why symmetric Hall bars show little rectification; and it is odd under exchange of the gate voltages,
\begin{equation}
\Delta R_\alpha(V_{\mathrm L},V_{\mathrm R})=-\Delta R_\alpha(V_{\mathrm R},V_{\mathrm L}),
\label{eq:gex}
\end{equation}
which is the observation of Ref.~\cite{Siddiki2010NJP} that reversing both the current direction and the lateral asymmetry leaves the traces unchanged while reversing only one of them does not.

We are careful about the status of this statement. Equation~\eqref{eq:DR2} proves one implication only: symmetric confinement implies vanishing rectification, so a measured $\Delta R_\alpha\neq0$ is a strong signature of broken inversion within the stated model and contact mapping. The converse does not hold. Broken inversion alone does not produce a finite $\Delta R_\alpha$: one also needs a nonlinear response, a current-induced electrostatic redistribution to carry the polarity information, and a non-zero operating current. $\Delta R_\alpha$ is therefore best described as a symmetry-allowed diagnostic --- a nonlinear response coefficient odd under edge exchange --- rather than as an order parameter in the strict sense.

\subsection{Reversal of the magnetic field}

Since $\RH$ changes sign with $B$ at a fixed pair of voltage probes, it is the magnitude that is compared in experiment. Writing $\Delta R_\alpha(B)=R_\alpha(B,+I)-R_\alpha(B,-I)$, Eq.~\eqref{eq:comp} gives
\begin{equation}
\begin{split}\Delta\RL(-B)&=-\Delta\RL(B),\\ \Delta|\RH|(-B)&=-\Delta|\RH|(B).\end{split}
\label{eq:Bodd}
\end{equation}
(For the signed $\RH$ the two minus signs cancel and $\Delta\RH$ is formally even in $B$; this is a statement about the sign convention for $\VH$, not a separate effect.)

It is tempting to read Eq.~\eqref{eq:comp} as the trivial statement that one may turn the sample over, since a $\pi$ rotation about $\hat{\bm x}$ takes $y\to-y$ and $B_z\to-B_z$. That reading is wrong. Such a rotation also reverses $\hat{\bm z}$, and a heterostructure is not invariant under it: the top gates, the donor layer, the two-dimensional system and the substrate are stacked in a definite order. Equation~\eqref{eq:comp} is instead a statement of in-plane covariance under transformed contacts and boundaries, valid within the model of Sec.~\ref{sec:model}, not a symmetry of the physical apparatus that would hold for any description of it.

Its value is as a test with a definite expected sign. Spurious polarity-dependent signals in a dc measurement --- thermovoltages at the probes, contact asymmetry along the bar, heating drift between the $+I$ and $-I$ sweeps --- need not respect Eq.~\eqref{eq:Bodd}, whereas rectification arising from the lateral confinement must within the ideal model. Verifying that $\Delta\RL$ and $\Delta|\RH|$ change sign on magnet reversal, and that $|I_c^+(B)|=|I_c^-(-B)|$ for the breakdown currents, therefore discriminates against those artefacts. It says more than that, and the positive statement is worth making. A contribution $g(B^2)p_xI_y$, even in $B$, is forbidden by no counting argument; what forbids it is $\Mop$ covariance of the medium, that is, the absence of intrinsic chirality and of spin--orbit coupling in Eq.~\eqref{eq:H}. Combined with the exactness of $\Pop$, oddness in $B$ is therefore the statement that the rectification is purely orbital, and the residual even-in-$B$ component is not merely an artefact one hopes is small but a measurement of the non-orbital contribution --- in a crystal that does lack inversion symmetry, as the Berry-curvature-dipole discussion of Sec.~\ref{sec:class} concedes. The qualification this test carries is narrower than the one on Eq.~\eqref{eq:gex}. By Eq.~\eqref{eq:Udis} the lateral impurity configuration is left untouched by the composed operation, so the residual even-in-$B$ component is set by contact asymmetry along the bar and by any $y$ dependence of the disorder, and not by the frozen lateral potential --- which is the dominant source of scatter in the gate-exchange test. Two practical qualifications belong here, since we are recommending Eq.~\eqref{eq:Bodd} partly on its robustness. The two traces are separated by a magnet sweep, and a gate-defined GaAs device is not always the same device afterwards: charge rearrangement in the doping layer and persistent photoconductivity both act between them. And at $\pm B$ the Hall probe pairs exchange roles, so probe placement along $y$ enters exactly where Eq.~\eqref{eq:Udis} locates the residual. Both are checkable by the probe permutation suggested below, which should be run alongside the field reversal rather than after it. It remains an ideal-model prediction rather than an unconditional null test, since contacts are never perfectly symmetric; but the expected residual is smaller, and its origin is different and independently checkable by permuting the probes.

We must be careful about what is and is not already known. Nonreciprocal transport in the quantum Hall regime of a gated conventional 2DEG has been observed: Iwakiri \emph{et al.}~\cite{Iwakiri2023} report a response quadratic in the current in a back-gated GaAs sample, depending on the out-of-plane field and, importantly, on the contact configuration, and attribute the inversion breaking to the back gate together with $B$. Giant nonreciprocal charge transport originating from quantum Hall states has also been reported in a topological insulator~\cite{Li2024}. What has not been done, to our knowledge, is the specific pair of tests that Eqs.~\eqref{eq:gex} and \eqref{eq:Bodd} prescribe on one and the same laterally asymmetric bar; the measurements of Ref.~\cite{Siddiki2010NJP} exchange the gate voltages but do not reverse the field, and the edge-to-bulk experiment of Ref.~\cite{Sirt2025APL}, while directly relevant background, addresses a different question.

\subsection{Distinguishing intrinsic rectification from gate and contact artefacts}\label{sec:artefacts}

Reference~\cite{Iwakiri2023} makes the need for a discriminating test concrete rather than hypothetical, since it establishes that a gate can by itself produce nonreciprocity in this regime. The symmetry analysis supplies the discriminator, and it is worth stating in the same language as Sec.~\ref{sec:class}.

A back gate breaks inversion along $\hat{\bm z}$, so the associated polar vector is $\bm q=q_z\hat{\bm z}$. With $\bm B=B_z\hat{\bm z}$ one has $\bm q\times\bm B=0$ and $\bm q\cdot\bm I=0$ for $\bm I\parallel\hat{\bm y}$: neither invariant of Sec.~\ref{sec:class} is available, which is consistent with the observation of Ref.~\cite{Iwakiri2023} that the effect there depends on which contacts are used, i.e.\ that the residual symmetry breaking is supplied by the contact geometry rather than by a bulk polar axis. A longitudinal gradient along the bar, whatever its origin, is described by a polar vector $\bm q\parallel\hat{\bm y}$, giving the invariant $\bm q\cdot\bm I=q_yI_y$; this is odd in $I$ and therefore contributes to $\Delta R_\alpha$, but it contains no factor of $B$ and hence is even under field reversal.

Two experimental separations follow, and they are independent:
\begin{enumerate}
\item \emph{Exchange of the lateral gate voltages, Eq.~\eqref{eq:gex}.} A back gate, a longitudinal gradient and a contact asymmetry are all unaffected by interchanging $V_{\mathrm L}$ and $V_{\mathrm R}$, so any $\Delta R_\alpha$ they produce is unchanged; confinement-driven rectification must change sign. This is the test already partially performed in Ref.~\cite{Siddiki2010NJP}. Its accuracy in a single device is limited by the frozen lateral disorder, which the gate exchange does not mirror.
\item \emph{Reversal of the field, Eq.~\eqref{eq:Bodd}.} Contributions of the form $q_yI_y$ are even in $B$; the confinement-driven term $B_zp_xI_y$ is odd. By Eq.~\eqref{eq:Udis} this test does not require the lateral disorder to transform, and it is correspondingly the more accurate of the two in a single device; what it does require is that the contacts map under source--drain exchange.
\end{enumerate}
An artefact would have to be odd under gate exchange \emph{and} odd in $B$ to survive both, which requires it to carry the same polar axis $\hat{\bm x}$ as the confinement itself. Concretely, we suggest the following protocol, all within one cooldown:
\begin{enumerate}
\item acquire full $R_\alpha(I)$ sweeps at both current polarities, at $\pm B$, and for both gate settings $(V_{\mathrm L},V_{\mathrm R})$ and $(V_{\mathrm R},V_{\mathrm L})$ --- eight traces;
\item interpolate every trace to a common $|I|$ grid, since fixed source--drain voltage does not give fixed current once $\RL$ differs between polarities, as Sec.~\ref{sec:mech} shows for the calculation itself;
\item form the current-odd component $\delta R=\tfrac12[R(+I)-R(-I)]$ for each of the four configurations;
\item decompose $\delta R$ into its parts even and odd under gate exchange and under field reversal. Only the doubly odd part is the confinement-driven response of Eq.~\eqref{eq:class}.
\end{enumerate}
The tolerance is set by the sample-specific fluctuation of the plateau edges rather than by instrumental noise, so the useful figure of merit is the ratio of the doubly odd component to the residual, not its absolute size. We regard the two tests together, on a single cooldown, as the decisive measurement. Criteria 1--3 of Sec.~\ref{sec:class} separate the confinement-driven term in addition from the Berry-curvature-dipole contribution, and the bounded field window of Sec.~\ref{sec:mech} separates it from the viscous and ballistic mechanisms discussed in Sec.~\ref{sec:hydro}.

Frozen disorder sets the accuracy with which either test can be expected to hold. As noted after Eq.~\eqref{eq:cov}, exchanging $V_{\mathrm L}$ and $V_{\mathrm R}$ does not mirror the impurity configuration of a given sample, so Eq.~\eqref{eq:gex} is exact only for the disorder-averaged model or between mirror-conjugate realisations. In a single device the antisymmetry will be violated at the level of the sample-specific fluctuation of the plateau edges, which is why the tests are best performed on the high-mobility, gate-defined bars of Ref.~\cite{Siddiki2010NJP}, and why agreement should be assessed as the dominance of the antisymmetric component rather than as an exact cancellation.

\subsection{Collapsing one strip: a test between the two current pictures}\label{sec:collapse}

The two tests above probe the symmetry of the response. A third measurement, of a different character, discriminates between the co-directed description used here and the chiral decomposition of Ref.~\cite{SirtLudwig2026}, and it uses the same samples and the same gates.

In a chiral description the imposed current at a given polarity resides on one edge, so at that polarity the width of the other strip is irrelevant. In the co-directed description of Eqs.~\eqref{eq:G}--\eqref{eq:RG} it is not, because the strips act as conductances in parallel and the wider one carries the current whichever edge it lies on. Making one edge progressively harder, so that its incompressible strip narrows towards collapse, therefore separates the two predictions: the chiral picture expects quantization to be lost at the polarity that selects the hard edge, whereas Eq.~\eqref{eq:RG} expects the surviving soft-edge strip to continue carrying the current, with the plateau retained at both polarities until that strip too is lost. The observable is the ratio $\Delta R_\alpha$ against the hard-edge gate voltage at fixed soft-edge voltage: the co-directed description predicts that it saturates once the hard strip has ceased to contribute, whereas the chiral description predicts a rapidly increasing rectification as the selected hard-edge strip approaches collapse. We put it this way rather than as a divergence: a real device enters breakdown, nonlocal transport, heating and contact redistribution well before any literal divergence could be reached, and the experimentally sufficient statement is growth against saturation. Because it is a saturation-versus-growth comparison rather than a comparison of magnitudes, it is insensitive to the model dependence discussed in Sec.~\ref{sec:conv}. We would welcome the measurement.

\section{Numerical evidence}

\subsection{Mechanism and computed profiles}\label{sec:mech}

The calculation reported in this section is intended to establish the mechanism and the transformation properties of the response, not to predict the absolute magnitude of the rectification or the precise field at which its window lies. Section~\ref{sec:conv} quantifies which of its outputs survive changes of the model parameters and which do not, and Sec.~\ref{sec:disorder} sets out one respect --- the position of the plateau relative to the classical Hall line --- in which the present disorder model is known to be inadequate. The statements we ask the reader to carry away from the numerics are the sign of the effect, its behaviour under the operations of Sec.~\ref{sec:ops}, and the existence of a bounded window in field. The spin-degenerate treatment used throughout, and what changes at odd $\nu$, are set out in Sec.~\ref{sec:notcover}.

That $\Pop$ is broken by an asymmetric $\Vconf$ is a statement about the configuration; making the breaking visible requires a mechanism coupling $\operatorname{sgn}I$ to the electrostatics, and in the screening theory this is the current-induced tilt of $\mu^*$. In the quantized regime essentially the whole Hall voltage drops across the ISs, so the local electrochemical potential at the two edges is displaced by $\pm e\VH/2$, with opposite signs for the two polarities~\cite{Siddiki2009EPL}. As intuition only, a Chklovskii-type estimate~\cite{Chklovskii1992} in which the available gap is shifted by the local drop of $\mu^*$,
\begin{equation}
\begin{gathered}W\simeq\left(\frac{2\bar\kappa\,\Delta E_{\mathrm{eff}}}{\pi^2e^2\,|d\nel/dx|}\right)^{1/2},\\ \Delta E_{\mathrm{eff}}=\hbar\omega_c\pm\tfrac12e\VH,\end{gathered}
\label{eq:W}
\end{equation}
suggests that the two edges respond with opposite signs. Equation~\eqref{eq:W} is a heuristic and is not used to establish anything; the strip widths quoted below come from the self-consistent solution.

\begin{figure*}[t]
\centering
\includegraphics[width=\textwidth]{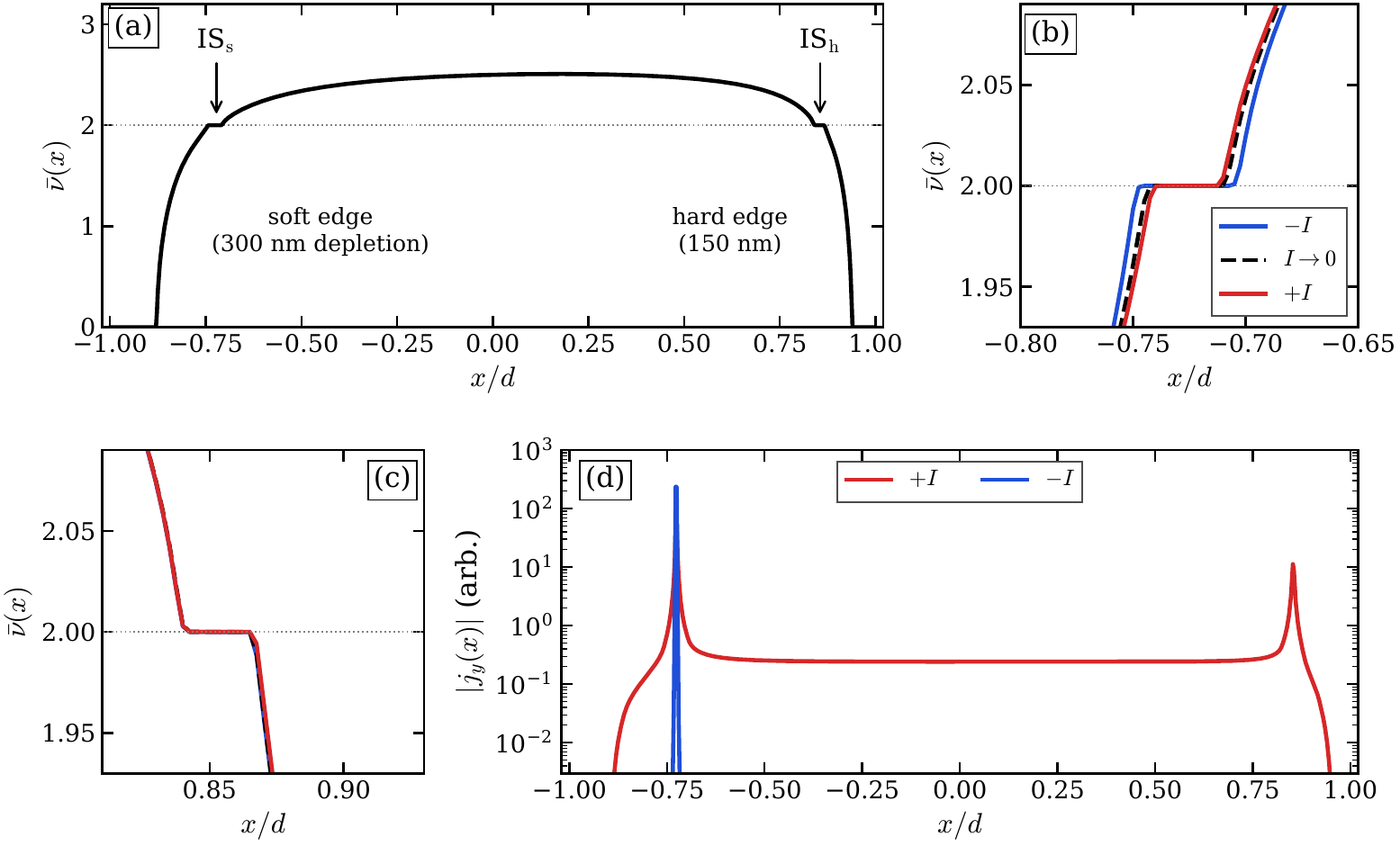}
\caption{Self-consistent solution for a hard--soft Hall bar ($d=2.5\,\mu$m, $n_0=4\times10^{11}$\,cm$^{-2}$, depletion 300\,nm left and 150\,nm right, $\nu(0)=2.5$, $kT=5\times10^{-3}\EFc$), obtained with the \scblin\ code~\cite{SCBLIN}. (a) Self-consistent filling-factor profile at $I\to0$, averaged over 40\,nm; the $\nu=2$ strips sit at the soft (IS$_{\rm s}$) and hard (IS$_{\rm h}$) edge. (b) Enlargement of the soft-edge strip and (c) of the hard-edge strip at the three currents of Table~\ref{tab:resp}, $-I$ (blue), $I\to0$ (black dashed) and $+I$ (red), compared at matched $|I|\simeq0.2\,\mu$A rather than at matched Hall voltage, so that the imposed $e\VH/\EFc$ is $-0.2000$ at $-I$ and $+0.1877$ at $+I$; the soft-edge strip measures 106, 81 and 69\,nm at $-I$, $I\to0$, $+I$, the hard-edge strip 56\,nm throughout. Mesh 800 intervals (6.25\,nm). (d) Current distribution $|j_y(x)|$ on a logarithmic scale. This is the imposed transport-current component of the quasi-local model, Eq.~\eqref{eq:j}, which by construction has one sign; the closed-loop persistent component of Refs.~\cite{SirtLudwig2026,Sirt2025NJP} carries no charge between the contacts and is not plotted.}
\label{fig:profile}
\end{figure*}

Figure~\ref{fig:profile} shows that solution, obtained by solving Eqs.~\eqref{eq:V}--\eqref{eq:R} together with SCBA spectral functions and conductivities and iterating the current-induced $\mu^*(x)$ to convergence~\cite{SCBLIN}. The bar has $d=2.5\,\mu$m, $n_0=4\times10^{11}$\,cm$^{-2}$, Gaussian scatterers of range $R=10$\,nm with $\Gamma/\hbar\omega_c=0.3$ at 10\,T, and depletion lengths of 300\,nm on the left (soft) and 150\,nm on the right (hard) edge. This broadening corresponds to a transport mobility $\mu\simeq4.2$\,m$^2$/Vs ($4.2\times10^4$\,cm$^2$/Vs) and a mean free path of about $0.4\,\mu$m in a $5\,\mu$m bar, well below that of the samples of Refs.~\cite{Siddiki2010NJP,Horas2008}: the uniform SCBA $\Gamma$ enters as a level width and is not a model of the disorder potential of those devices, and Sec.~\ref{sec:disorder} sets out what that omission costs. At $\nu(0)=2.5$, $kT=5\times10^{-3}\EFc$ and $e\VH=\pm0.2\EFc=\pm2.6$\,meV ($I\simeq0.2\,\mu$A), the soft-edge strip measures 106, 81 and 69\,nm at $-I$, $I\to0$ and $+I$, while the hard-edge strip stays at 56\,nm. The transport consequence is collected in Table~\ref{tab:resp} and visible in the current distribution, Fig.~\ref{fig:profile}(d). At $-I$ the current is carried by the wide soft-edge strip and the response is quantized to six digits; at $+I$, at the same $|I|$, the strip can no longer carry it, the current spreads over the compressible bulk, and $\RH$ falls 6\% below the plateau while $\RL$ rises by four orders of magnitude. The two polarities straddle the zero-current result. The field has deliberately been chosen near the low-field end of the plateau, where the zero-current response is itself only marginally quantized: this is the regime in which rectification lives, and deep inside a plateau both polarities are quantized and $\Delta R_\alpha$ vanishes.

\begin{table*}[t]
\centering
\caption{Computed response of the hard--soft bar at $\nu(0)=2.5$, i.e.\ $\hbar\omega_c=10.34$\,meV and $B=5.98$\,T, for the three currents of Fig.~\ref{fig:profile}, compared at matched $|I|$ rather than at matched $\VH$; the imposed Hall voltage needed to reach that current is listed. This is the operating point at which the nominal coefficient $\gamma$ of Sec.~\ref{sec:class} is evaluated. $W_{\mathrm s}$ and $W_{\mathrm h}$ are the widths of the soft- and hard-edge $\nu=2$ strips. Mesh $N=800$ (6.25\,nm).}
\label{tab:resp}
\small\setlength{\tabcolsep}{4pt}
\begin{tabular}{lcccccc}
\toprule
current & $I$ ($\mu$A) & $e\VH/\EFc$ & $W_{\mathrm s}$ (nm) & $W_{\mathrm h}$ (nm) & $\RH$ ($h/2e^2$) & $\RL$ ($h/e^2$)\\
\midrule
$-I$ & $-0.2004$ & $-0.2000$ & 106 & 56 & 0.999994 & $1.2\times10^{-6}$\\
$I\to0$ & $0.0051$ & $0.0050$ & 81 & 56 & 0.9866 & $2.6\times10^{-3}$\\
$+I$ & $0.2001$ & $0.1877$ & 69 & 56 & 0.9396 & $1.2\times10^{-2}$\\
\bottomrule
\end{tabular}
\end{table*}

The calculation is not fitted to the data of Refs.~\cite{Siddiki2009EPLb,Siddiki2010NJP}, but it is not parameter-free either: it takes as input the two depletion lengths, the donor density, the temperature, the impurity range and strength, the averaging length $\lambda$, the mesh, and the imposed Hall voltage. All of these are specified above, and their influence is quantified next.

\begin{figure*}[t]
\centering
\includegraphics[width=\textwidth]{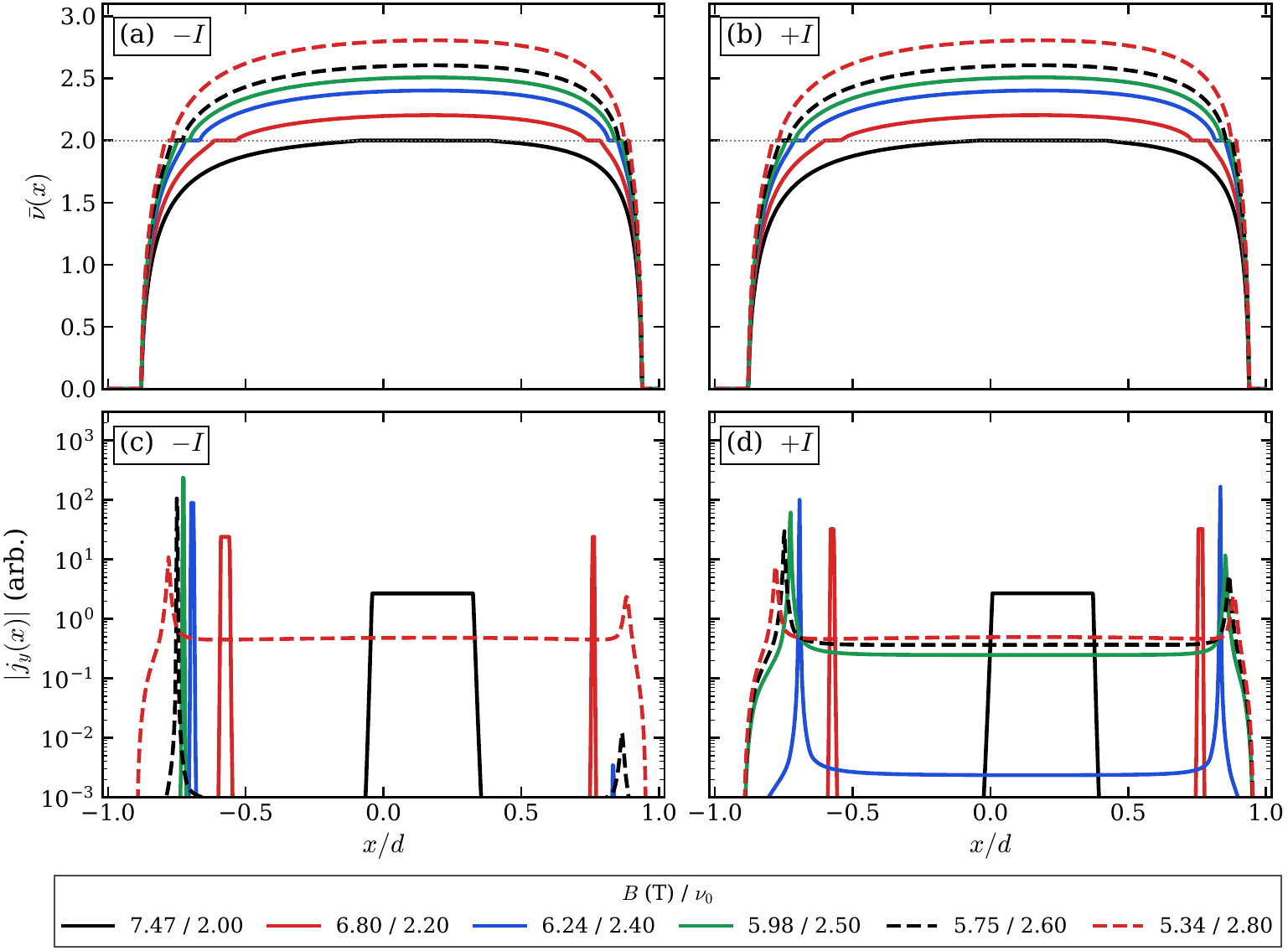}
\caption{Calculated filling factor and current density of the hard--soft bar ($300|150$\,nm) at six magnetic fields spanning the $\nu_0=2$ plateau. (a) $\bar\nu(x)$ at $-I$ and (b) at $+I$, averaged over $\lambda=40$\,nm; the dotted line marks $\nu=2$, and each flat section on it is an incompressible strip. (c),(d) the corresponding current density $|j_y(x)|$ on a logarithmic scale, same fields, at $-I$ and $+I$ respectively. All at matched $|e\VH|=0.2\EFc$; every curve is a separate self-consistent solution of Eqs.~\eqref{eq:V}--\eqref{eq:R} ($kT=5\times10^{-3}\EFc$, $N=800$), and the legend gives $B$ with the corresponding $\nu_0$ (solid black, red, blue, green, then dashed black, red, in order of decreasing field). At $\nu_0=2.0$ the incompressible region occupies the middle of the bar and carries the current there; as the field falls it breaks into two edge strips which narrow and move outward. The two polarities coincide until the strips are narrow enough for one of them to fail, at $\nu_0\simeq2.4$, and very nearly coincide beyond the plateau at $\nu_0=2.8$; on the scale of (a) and (b) the polarity dependence of the strip widths is not resolved, and Fig.~\ref{fig:profile}(b),(c) show it magnified at one field. The computed responses are $\RH=1.000000$, 1.000000, 0.999998, 0.999994, 0.999738, 0.812154 at $-I$ and 1.000000, 1.000000, 0.999551, 0.938379, 0.889223, 0.806954 at $+I$, in units of $h/2e^2$ and in order of decreasing field. As in Fig.~\ref{fig:profile}(d), this is the imposed transport-current component of the quasi-local model, which by construction has one sign.}
\label{fig:sweep}
\end{figure*}

The single field of Table~\ref{tab:resp} is not representative, and it is worth seeing what the calculation gives across the plateau. Figure~\ref{fig:sweep} collects six fields spanning $\nu_0=2.0$ to 2.8 at both polarities and at the same $|e\VH|=0.2\EFc$, showing the filling factor in panels (a),(b) and the current density it produces in (c),(d); every curve is a separate converged solution.

The density panels give the electrostatics on its own, before any transport enters. At the high-field end, $\nu_0=2.0$, the profile is pinned at $\nu=2$ across almost the entire bar: the incompressible region is the bulk, and the two narrow compressible flanks carry no weight. Lowering the field raises $\nu$ in the centre above 2, the pinned region parts in the middle, and what survives is a pair of strips that migrate towards the edges and narrow as they go --- at $\nu_0=2.2$ they are centred at $|x|/d=0.57$ and 0.76 and are 194 and 119\,nm wide, by $\nu_0=2.6$ they have moved to 0.75 and 0.87 and narrowed to 94 and 44\,nm, and at $\nu_0=2.8$ the flat sections have gone altogether. The two edges are not equivalent at any of these fields, the soft-edge strip being the wider throughout, which is the broken $\Pop$ of Sec.~\ref{sec:ops} expressed in the density. What the panels do \emph{not} resolve is the polarity: (a) and (b) are indistinguishable by eye, because the current-induced change in strip width is at most a few tens of nanometres against a $5\,\mu$m bar, and the magnified view of Fig.~\ref{fig:profile}(b),(c) is needed to see it.

The current panels show why that sub-percent change nevertheless decides the response. The weight $1/\rho_l$ in Eq.~\eqref{eq:R} is exponentially large where $\nu$ is pinned, so the transport reads the density profile through an exponential magnifier: a strip either carries essentially all of the current or none of it, with little in between. Three regimes follow, ordered in field. Deep inside the plateau, at $\nu_0=2.0$, the current is carried in the middle of the bar --- the edge-to-bulk migration within a single plateau that was recently resolved with a central ohmic probe~\cite{Sirt2025APL} --- and the two polarities are indistinguishable, $\RH$ being quantized to nine digits. As the strips form and move outward the current follows them, and the response stays quantized at both polarities down to $\nu_0\simeq2.2$. Only when the strips have narrowed enough that one of them can no longer carry the imposed current does the polarity matter: at $\nu_0=2.4$ the $+I$ solution has begun to leak into the compressible bulk while the $-I$ solution has not, and at $\nu_0=2.5$ and 2.6 the separation is complete, the $+I$ current running through the bulk at a density three orders of magnitude above its $-I$ counterpart. Beyond the plateau, at $\nu_0=2.8$, neither polarity has a strip left and the distributions very nearly coincide again.

This is Eq.~\eqref{eq:RG} in the form of a picture. What distinguishes the polarities is not the presence of the structural asymmetry, which is there at every field in panels (a) and (b), but whether the wider strip can still carry the current. Rectification is accordingly a plateau-edge phenomenon with a finite window in field, bounded at high field by the robustness of the wide bulk incompressible region and at low field by the collapse of the wider strip. The two ends are not alike. At the high-field end it vanishes outright, $\RH=1.000000$ at both polarities for $\nu_0=2.0$ and 2.2. At the low-field end it falls by more than an order of magnitude but not to zero: at $\nu_0=2.8$, with no strip left at either polarity, $\Delta\RH=0.005$ survives against 0.11 at $\nu_0=2.6$, of the same size as the beyond-plateau row of Table~\ref{tab:lambda}. We attribute this to the compressible bulk, which retains the asymmetry of the underlying density profile once the strips have gone. That attribution is testable, and we have tested it: holding the field at $\nu_0=2.798$ and reducing the depletion asymmetry, the residual falls from $\Delta\RH=0.0052$ at $300|150$\,nm to 0.0016 at $225|150$\,nm and to $-1\times10^{-5}$ at $150|150$\,nm. It vanishes with the structural asymmetry, so it is a property of the confinement and not a numerical floor; the falloff is steeper than linear in $d_{\mathrm L}-d_{\mathrm R}$, as it must be if it is fed by the same current-induced redistribution that drives the effect inside the window. The loss of the strips therefore weakens the rectification without abolishing it. That is why we put forward the existence and the sign of the window, rather than the peak value inside it, as the prediction of Sec.~\ref{sec:conv}: the two bounds are set by the electrostatics of panels (a),(b), which is robust, whereas the peak value depends on where within the window a given field happens to fall.

\begin{figure*}[t]
\centering
\includegraphics[width=\textwidth]{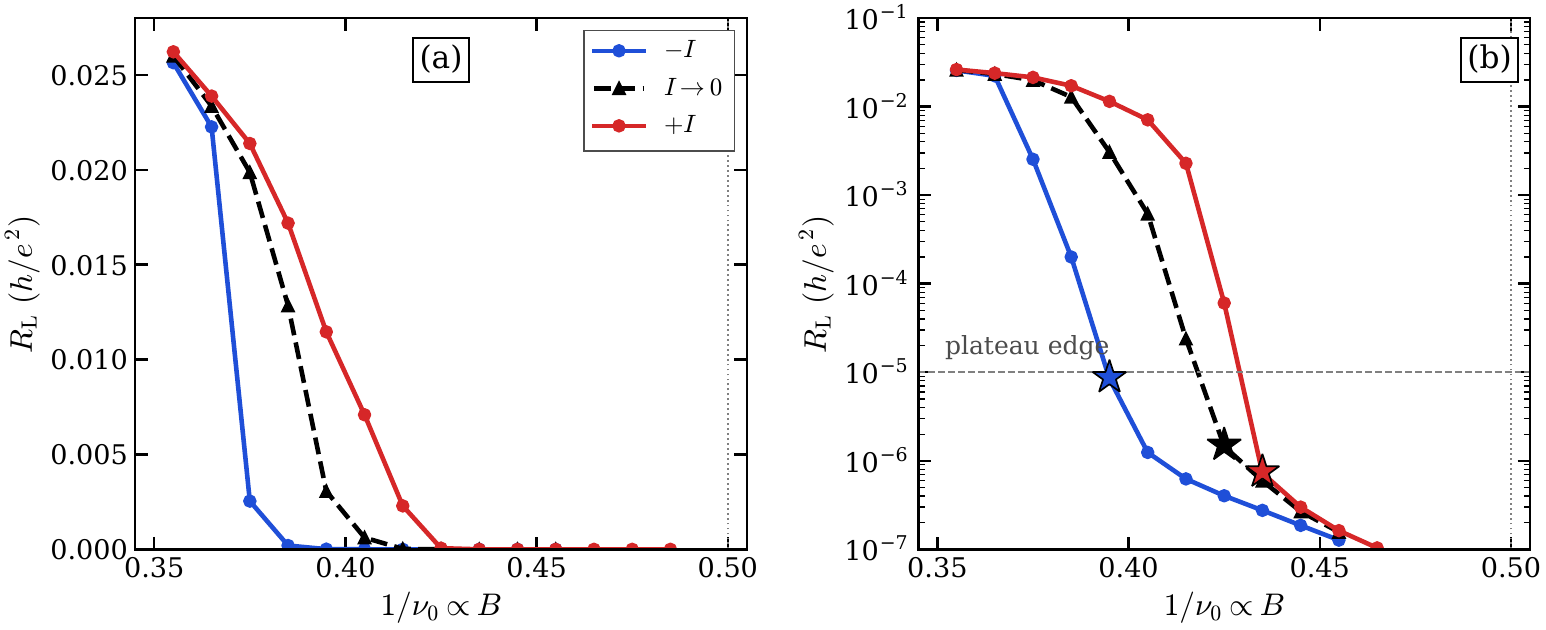}
\caption{Longitudinal resistance across the $\nu_0=2$ plateau of the hard--soft bar at $|I|\simeq0.15\,\mu$A, at $-I$ (blue), $+I$ (red) and in the linear-response limit $I\to0$ (black dashed), on (a) linear and (b) logarithmic scale. Stars in (b) mark where each trace first reaches $\RL=10^{-5}h/e^2$ (dashed horizontal line): the plateau edge moves with the current polarity, to $\nu_0=2.53$, 2.35 and 2.30 for $-I$, $I\to0$ and $+I$. The dotted vertical line is the classical crossing at $1/\nu_0=0.5$. The comparison here is at matched current; Fig.~\ref{fig:asym}(b) compares instead at matched imposed Hall voltage, which is why the quoted plateau edges differ.}
\label{fig:RL}
\end{figure*}

\begin{figure*}[t]
\centering
\includegraphics[width=\textwidth]{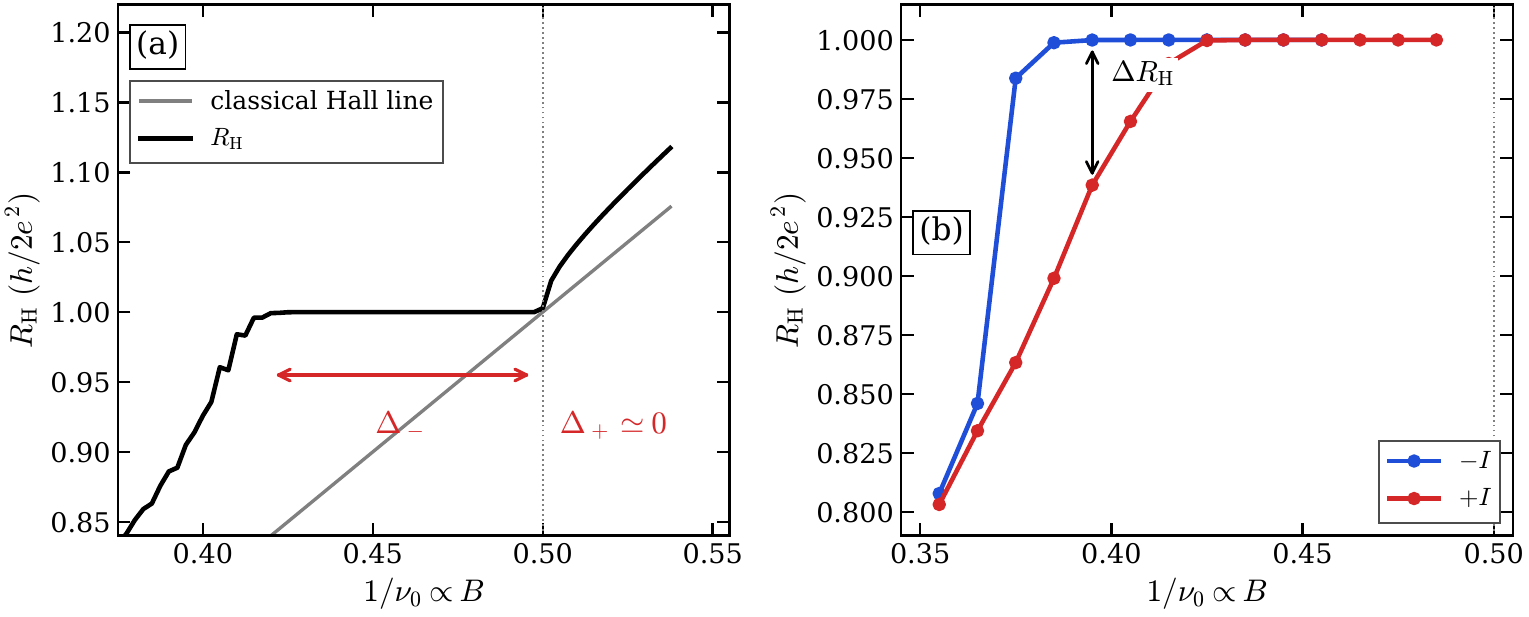}
\caption{(a) Plateau asymmetry in a symmetric-bar screening calculation, at small current: the computed $\nu_0=2$ plateau occupies $0.420\le1/\nu_0\le0.4975$, almost entirely on the low-field side of its intersection with the classical Hall line (grey) at $1/\nu_0=0.5$. The panel is shown for the one-sidedness of the asymmetry, not for the position of the edges: those are quoted in Sec.~\ref{sec:disorder} from the sweep of Fig.~\ref{fig:temp}, whose edge criterion $|\RH-1|<10^{-3}$ is stated and whose imposed current differs from the one used here. As discussed in Sec.~\ref{sec:disorder} this over-states the measured asymmetry and reflects the neglect of long-range disorder. (b) Rectification of the hard--soft bar at $e\VH=\pm0.15\EFc$, $-I$ (blue) and $+I$ (red): the plateau survives to $\nu_0=2.60$ for one polarity and only to 2.35 for the other. By Eq.~\eqref{eq:comp} the two curves must interchange when $B$ is reversed.}
\label{fig:asym}
\end{figure*}

Figure~\ref{fig:RL} extends this from the single field of Table~\ref{tab:resp} to the whole $\nu_0=2$ plateau, showing $\RL$ for the two polarities and in the linear-response limit. The three traces are ordered throughout: taking $\RL<10^{-5}h/e^2$ as the criterion, the plateau extends to $\nu_0=2.53$ at $-I$, to 2.35 at $I\to0$ and only to 2.30 at $+I$. The rectification is therefore a displacement of the low-field plateau edge. The zero-current trace lies between the two polarities rather than coinciding with either --- as Eq.~\eqref{eq:cov} requires, since $\Delta R_\alpha$ is a difference between mirror-conjugate devices and not a deviation from the $I=0$ result --- and the effect is visible over a range of field rather than at one point, which is what makes it accessible to the tests of Sec.~\ref{sec:artefacts}. Figure~\ref{fig:asym}(b) shows the same displacement in $\RH$, compared there at matched imposed Hall voltage $e\VH=\pm0.15\EFc$ rather than at matched current: the plateau survives to $\nu_0=2.60$ at one polarity and only to 2.35 at the other, and by Eq.~\eqref{eq:comp} the two traces must interchange when the field is reversed.

\subsection{Parallel-conductance structure}

Because $j_y$ has one sign in this prescription, the integrals in Eq.~\eqref{eq:R} split into a sum of positive contributions,
\begin{equation}
I=E^0_y\sum_iG_i,\qquad G_i=\int_i\frac{dx}{\rho_l(x)}>0,
\label{eq:G}
\end{equation}
so the strips act as conductances in parallel and
\begin{equation}
\RL=\frac{2d}{\sum_iG_i},\qquad \RH=\frac{\sum_iG_i\,\rho_{\mathrm H}^{(i)}}{\sum_iG_i},
\label{eq:RG}
\end{equation}
i.e.\ $\RH$ is a conductance-weighted average of the local Hall resistivity. If every region with appreciable weight has $\nu=N$ then $\rho_{\mathrm H}^{(i)}=h/Ne^2$ for all of them and $\RH$ is exactly quantized independently of the widths, which is why the plateau value is insensitive to the confinement while its extent is not. Quantization is lost only when the compressible bulk acquires weight --- precisely what Fig.~\ref{fig:profile}(d) shows at $+I$ and not at $-I$.

Unidirectionality is not what makes rectification possible: the chiral alternative of Ref.~\cite{SirtLudwig2026} would also give a polarity-dependent response, and arguably a larger one. Nor is Eq.~\eqref{eq:G} independent physics, since $\rho_l>0$ in any local dissipative model. What it determines is the \emph{form} of the answer: because the strips add, the response is governed by the weighted average~\eqref{eq:RG} and hence by whichever strip is widest, whereas in a chiral description it would be governed by one edge at a time and $\Delta R_\alpha$ would be a different function of the two widths. The two descriptions are therefore distinguishable in principle by the dependence of the rectification on the confinement asymmetry.

\subsection{Convergence and sensitivity}\label{sec:conv}

Table~\ref{tab:sens} collects the response at $\nu(0)=2.5$ and $e\VH=\pm0.2\EFc$ as the numerical and model inputs are varied one at a time.

The averaging length $\lambda$ deserves separate comment, since the table shows the rectification vanishing at $\lambda=20$\,nm.

First, $\lambda$ is not adjustable at will. It stands for the two microscopic lengths that the quasi-local prescription suppresses. From below it must exceed the extent of the wave functions, i.e.\ the magnetic length $\ell_B=25.7\,\mathrm{nm}/\sqrt{B[\mathrm T]}$, which is 11.0\,nm at 5.4\,T and 10.0\,nm at 6.6\,T: an average over a window narrower than $\ell_B$ retains strips that the wave functions would bridge. From above it should not exceed the scale on which a local conductivity can be defined, the mean electron spacing, i.e.\ the Fermi wavelength. Evaluated with the mean electron density of the bar, $\bar n_{\mathrm{el}}\simeq3\times10^{11}$\,cm$^{-2}$ rather than the donor density, $k_{\mathrm F}=(2\pi\bar n_{\mathrm{el}})^{1/2}=0.137$\,nm$^{-1}$ and $\lambda_{\mathrm F}=46$\,nm. The admissible window is therefore $\ell_B\lesssim\lambda\lesssim\lambda_{\mathrm F}$, that is 10--46\,nm here. Reference~\cite{Siddiki2004PRB} prescribes $\lambda\sim\lambda_{\mathrm F}/2\simeq23$\,nm and in practice uses 20--40\,nm, with 30\,nm as its standard value. We have used 40\,nm, at the upper end of that range; Table~\ref{tab:sens} brackets it with 20\,nm, near the lower end, and 60\,nm, outside the range altogether. So the parameter is bounded by the sample, not chosen freely, but we do not claim it is determined to better than a factor of two, and the reader should read the computed magnitudes with that latitude.

Second, and more important, $\lambda$ does not switch the mechanism on and off. What it sets is the width below which a strip stops carrying current, and hence the location of the plateau edge; since rectification is a plateau-edge phenomenon (Sec.~\ref{sec:mech}), changing $\lambda$ at fixed $B$ moves one into or out of the window in which the effect is visible. Repeating the calculation at each $\lambda$ and searching for that model's own plateau edge confirms this, Table~\ref{tab:lambda}: at $\lambda=20$\,nm the rectification reappears at lower field, $\nu_0=2.86$, where it is in fact larger than in the reference case ($\Delta\RH=0.23$ against 0.062); at $\lambda=60$\,nm quantization for both polarities is recovered at higher field, $\nu_0=2.20$. The apparent extreme sensitivity in Table~\ref{tab:sens} is an artefact of comparing different $\lambda$ at one fixed $B$. The core mechanism --- current-induced redistribution of strip widths under broken in-plane inversion --- is untouched by $\lambda$; only the field at which it is most visible moves.

\begin{table}[t]
\centering
\caption{The averaging length relocates rather than removes the effect. For each $\lambda$ the response is evaluated at a field near that model's own plateau edge; the last column states whether the operating point lies inside the plateau, at its edge, or beyond it.}
\label{tab:lambda}
\small\setlength{\tabcolsep}{3.5pt}
\begin{tabular}{cccccl}
\toprule
$\lambda$ (nm) & $\nu_0$ & $\RH(-I)$ & $\RH(+I)$ & $\Delta\RH$ & regime\\
\midrule
20 & 2.50 & 1.00000 & 1.00000 & 0.000 & inside\\
20 & 2.86 & 0.99999 & 0.772 & 0.228 & edge\\
40 & 2.50 & 0.999994 & 0.938 & 0.062 & edge\\
60 & 2.20 & 1.00000 & 0.99996 & 0.000 & inside\\
60 & 2.50 & 0.902 & 0.897 & 0.005 & beyond\\
\bottomrule
\end{tabular}
\end{table}

Two conclusions follow from the two tables. The sign of the effect --- $-I$ more strongly quantized than $+I$ for a soft-left bar --- is robust against every variation tried, as is its insensitivity to mesh refinement and to a 10\% change of both depletion lengths. The magnitude at a given field is not a robust prediction, because it depends on where the plateau edge lies, which in turn depends on $\lambda$, on the disorder broadening and on temperature. Quantitatively, the sign is unchanged in every row of Tables~\ref{tab:lambda} and~\ref{tab:sens}; evaluated at each model's own plateau edge the rectification lies between $\Delta\RH=0.06$ and 0.23, a factor of four across the admissible range of $\lambda$, and it falls to zero only when the operating point is moved inside the plateau. We therefore put forward the sign, the gate-exchange antisymmetry~\eqref{eq:gex} and the field antisymmetry~\eqref{eq:Bodd} as the predictions, and treat the magnitude at a given $B$ as model-dependent.

A separate check concerns Eq.~\eqref{eq:cov} itself. Solving the full nonlinear problem for the device $[300|150]$\,nm at $+I$ and, independently, for its mirror $[150|300]$\,nm at $-I$ gives
\begin{equation}
\max_x\bigl|\nu(x)-\tilde\nu(-x)\bigr|=5\times10^{-10},
\label{eq:mirrortest}
\end{equation}
with $\RH$ and $\RL$ agreeing to eight significant digits. This verifies that the implementation is covariant, as the equations require; it is a test of the code and not of the physical approximation, and we do not present it as evidence for the model.

\begin{table*}[t]
\centering
\caption{Sensitivity of the computed response at $\nu(0)=2.5$, $e\VH=\pm0.2\EFc$, in units of $h/2e^2$ for $\RH$ and $h/e^2$ for $\RL$. Reference row: mesh $N=800$, $\lambda=40$\,nm, $\gamma_I=0.3$, $kT=5\times10^{-3}\EFc$, depletion $300|150$\,nm. Each subsequent row changes one input.}
\label{tab:sens}
\small\setlength{\tabcolsep}{4pt}
\begin{tabular}{lccccc}
\toprule
 & $\RH(-I)$ & $\RL(-I)$ & $\RH(+I)$ & $\RL(+I)$ & $\Delta\RH$\\
\midrule
reference & 0.999994 & $1.2\times10^{-6}$ & 0.938 & $1.2\times10^{-2}$ & 0.062\\
mesh $N=400$ & 0.999994 & $1.1\times10^{-6}$ & 0.932 & $1.3\times10^{-2}$ & 0.068\\
$\lambda=20$\,nm & 1.00000 & $2.3\times10^{-7}$ & 1.00000 & $3.6\times10^{-7}$ & 0.000\\
$\lambda=60$\,nm & 0.902 & $2.0\times10^{-2}$ & 0.897 & $2.1\times10^{-2}$ & 0.005\\
$\gamma_I=0.2$ & 1.00000 & $4.9\times10^{-7}$ & 0.993 & $1.3\times10^{-3}$ & 0.007\\
$\gamma_I=0.4$ & 0.99926 & $1.5\times10^{-4}$ & 0.907 & $1.9\times10^{-2}$ & 0.092\\
$kT=10^{-2}\EFc$ & 0.99922 & $1.5\times10^{-4}$ & 0.933 & $1.3\times10^{-2}$ & 0.066\\
depletion $330|165$\,nm & 0.99999 & $1.0\times10^{-6}$ & 0.939 & $1.3\times10^{-2}$ & 0.061\\
\bottomrule
\end{tabular}
\end{table*}

\section{Discussion}

\subsection{Relation to the edge-channel and persistent-current pictures}\label{sec:LB}

The ideal $N$-channel LB response~\cite{Buttiker1988} is independent of $\Vconf$ and of $\operatorname{sgn}I$, and predicts $\Delta R_\alpha\equiv0$ even for a strongly asymmetric device. It is not in conflict with Eq.~\eqref{eq:cov} --- it satisfies it trivially --- but it possesses an accidental invariance, independence of the confinement, which the Hamiltonian~\eqref{eq:H} does not have. What a measured $\Delta R_\alpha\neq0$ would falsify is that accidental invariance --- the ideal, confinement-independent limit --- and not Landauer--B\"uttiker theory as a framework, nor the existence of chirality. We make no claim to either. A finite-bias treatment with self-consistent electrostatics, energy-dependent transmission and realistic contacts can perfectly well acquire a dependence on the confinement; what it cannot do is retain the ideal perfectly-transmitted-channel result and still produce $\Delta R_\alpha\neq0$. That the current path moves with $B$ within a single plateau~\cite{Sirt2025APL} is by itself outside that description, in which it cannot move at all.

Equation~\eqref{eq:cov} is silent, by construction, on the decomposition of the total current density, and we have been careful above to keep it that way. It is worth setting out none the less why we retain the co-directed transport current of the quasi-local model rather than the chiral decomposition of Ref.~\cite{SirtLudwig2026}, and where we see a tension that we think has not yet been resolved.

The first point is internal to the screening theory. The account of the quantized plateaus given in Refs.~\cite{Guven2003,Siddiki2004PRB} --- finite plateau widths, exactly quantized values, and the transition from edge to bulk current within a plateau --- is obtained from the local Ohm's law~\eqref{eq:ohm} and the singular integrals of Eq.~\eqref{eq:R}. Within that relation $j_y=E^0_y/\rho_l$ with $\rho_l>0$, so a counter-propagating transport component is not a small correction but is inexpressible: the local Ohm's law and a chiral imposed current cannot both be kept. Adopting the chiral decomposition therefore means relinquishing the framework from which the plateau structure was derived in the first place, and it is not yet clear to us what replaces it. We say this as a question rather than an objection.

The second concerns the role assigned to the persistent current. As Ref.~\cite{SirtLudwig2026} itself emphasises, $\bm j_{\mathrm{pers}}$ is an equilibrium quantity that closes inside the sample and transports no charge between the contacts. That it should nevertheless determine where the imposed current flows is an additional dynamical assumption, not a consequence of its existence; in a linear local response the distribution of $\bm j_{\mathrm{impo}}$ is fixed by $\hat\rho(x)$ and the driving field, and the equilibrium circulating current drops out.

The third is what the asymmetry measurements actually constrain. The rectification of Refs.~\cite{Siddiki2009EPL,Siddiki2010NJP} is governed by the slope of the confinement potential: hard and soft edges enter through $|d\nel/dx|$, hence through the strip widths, and interchanging them interchanges the two polarities. Both descriptions reproduce that interchange, so the existing data do not by themselves settle the question --- but they are naturally accommodated by Eqs.~\eqref{eq:G}--\eqref{eq:RG}, in which the two strips act as conductances in parallel and the response is set by whichever is wider, and they require no additional ingredient there.

That last observation supplies a third experiment, distinct in character from the two symmetry tests because it discriminates between the two current pictures rather than between mechanisms of nonreciprocity. It is set out in Sec.~\ref{sec:collapse}.

\subsection{Symmetry classification of the rectification}\label{sec:class}

It is natural to ask where this effect sits among known nonreciprocal responses. The familiar phenomenology of electrical magnetochiral anisotropy~\cite{Rikken2001,Rikken2005},
\begin{equation}
R=R_0\bigl(1+\gamma\,\bm B\cdot\bm I\bigr),
\label{eq:MCA}
\end{equation}
does not apply here: in the Hall-bar geometry $\bm B=B_z\hat{\bm z}$ and $\bm I=I_y\hat{\bm y}$, so $\bm B\cdot\bm I=0$ identically and the inner-product invariant vanishes. The effect discussed in this paper is therefore not conventional MCA, and should not be called so.

The correct classification uses the structural polar vector of the asymmetric confinement. Let $\bm p=p_x\hat{\bm x}$ characterise the lateral asymmetry, with $p_x\propto V_{\mathrm L}-V_{\mathrm R}$ to leading order, odd under $x\to-x$ and vanishing for symmetric gates. The lowest-order scalar that can be built from $\bm p$, $\bm B$ and $\bm I$ and that is odd in the current is of vector-product type,
\begin{equation}
\delta R\propto B_z(\bm p\times\bm I)_z=B_zp_xI_y,
\label{eq:class}
\end{equation}
equivalently $R=R_0[1+\gamma(\bm B\times\bm p)\cdot\bm I]$, where $\delta R$ denotes the component of the response odd in the current, $\delta R=\tfrac12[R(+I)-R(-I)]=\tfrac12\Delta R_\alpha$. Equation~\eqref{eq:class} is a lowest-order classification: it presumes an analytic weak-current expansion about a chosen operating state, and it fixes the transformation parity of the odd component, not its magnitude or its linearity in $I$. The computed plateau-edge response is strongly nonlinear and of order unity, so Eq.~\eqref{eq:class} must not be read as a fit formula for the full traces. With that understood, Eq.~\eqref{eq:class} reproduces both relations derived above without further input: it is odd under $p_x\to-p_x$, which is Eq.~\eqref{eq:gex}; odd under $I\to-I$, so that it contributes to $\Delta R_\alpha$ and not to the reciprocal part; and odd under $B_z\to-B_z$, which is Eq.~\eqref{eq:Bodd}. It is invariant under the full $\Pop$, which reverses $p_x$ and $I_y$ together --- as it must be, $\Pop$ relating the device to its mirror partner rather than to itself. We therefore describe the effect as a gate-tunable, magnetic-field-induced nonreciprocal response, or magneto-asymmetric rectification, and reserve the term MCA for the inner-product case.

\paragraph{Magnitude in standard metrics} For comparison with the nonreciprocal-transport literature we quote three figures at the operating point of Table~\ref{tab:resp} ($\nu(0)=2.5$, $B=5.98$\,T, $|I|=0.200\,\mu$A), where $\Delta\RH/\RH=0.062$. The nonreciprocity ratio is $\delta R/R_0=\tfrac12\Delta\RH/\RH=3.1\times10^{-2}$. The second-order Hall coefficient, the current-odd part of the Hall resistance per unit current, is
\begin{equation}
\frac{\delta\RH}{|I|}=\frac{\Delta\RH}{2|I|}\simeq2.0\times10^{9}\ \Omega\,\mathrm A^{-1}.
\label{eq:coef}
\end{equation}
The nominal magnetochiral-type coefficient, defined through $\delta R/R_0=\tfrac12\gamma B_zI_y$, is $\gamma\simeq5\times10^4$\,A$^{-1}$T$^{-1}$, comparable to the largest values reported for nonreciprocal transport in the quantum Hall regime~\cite{Li2024} and several orders of magnitude above typical bulk noncentrosymmetric conductors~\cite{Rikken2001,Ideue2017}; surveys of reported magnitudes across mechanisms are given in Refs.~\cite{Du2021,SuarezRodriguez2025}.

All three are local figures of merit at a stated operating point and not material constants. The response is not linear in $I$ at the plateau edge, being of order unity there and vanishing deep inside the plateau, so none of these numbers should be carried into a table of material coefficients, and each would change by orders of magnitude if evaluated at $\nu_0=2.0$. The nonreciprocity number of Ref.~\cite{Kirkinis2025}, defined for a viscous liquid through the velocity dependence of the shear viscosity, has no counterpart here, for the reason given in Sec.~\ref{sec:hydro}: the system is not in the hydrodynamic regime.

\paragraph{Relation to the second-order conductivity tensor} It is worth saying where Eqs.~\eqref{eq:cov} and~\eqref{eq:class} sit within the general phenomenology of second-order response, since the two are easily conflated. Writing the current-odd part of the field as $E^{(2)}_a=\rho_{abc}\,j_bj_c$ with a rank-three tensor $\rho_{abc}(B)$ of the medium, the two observables used here are $\rho_{xyy}$, which carries the current-odd part of the Hall field, and $\rho_{yyy}$, which carries that of the longitudinal field, since $\bm j\parallel\hat{\bm y}$. Equation~\eqref{eq:class} is then the statement that both components are, to leading order, proportional to $B_zp_x$: forbidden in a mirror-symmetric device, first order in the structural polar vector, and odd in the field. In that sense the classification is a special case of the general rank-three framework and we do not claim otherwise; its content is that $\bm p$ is here a gate-tunable quantity rather than a lattice one.

Equation~\eqref{eq:cov} is not a special case of that framework. A tensor expansion presumes analyticity in the current about a chosen operating state and assigns the response to a finite set of coefficients. The covariance relation is a property of the full nonlinear self-consistent problem: it holds where the response is of order unity and not expandable --- as it is at the plateau edge, where $\Delta\RH/\RH=0.062$ and $\RL$ differs between the polarities by four orders of magnitude --- and it constrains the exact solution rather than the leading coefficients. The relation between the two is that Eq.~\eqref{eq:cov} implies the parity assignments of Eq.~\eqref{eq:class} wherever the expansion exists, and continues to constrain the response where it does not.

Two features distinguish it from the material realisations of either type~\cite{Ideue2017,Wakatsuki2017,Du2021}. The inversion breaking is not intrinsic: in a polar crystal it is a fixed property of the lattice, whereas here it is imposed by the gates, so it is continuously tunable, reversible \emph{in situ} and can be switched off, which is what makes Eq.~\eqref{eq:gex} a controlled test rather than a comparison between samples. And the mechanism is not relativistic: nonreciprocity in bulk conductors is usually traced to spin--orbit coupling and is correspondingly small, whereas here it arises from the current-induced rearrangement of a self-consistent electrostatic state and is of order unity.

\paragraph{Relation to the Berry-curvature-dipole nonlinear Hall effect} The Berry-curvature-dipole (BCD) mechanism~\cite{Sodemann2015,Ortix2021,Du2021,SuarezRodriguez2025} occupies the remaining corner of the same symmetry table: $\Top$ intact, inversion broken by the crystal, and a second-order Hall current governed by a band-geometric Fermi-surface quantity. Since GaAs is a zincblende crystal and therefore lacks a centre of inversion, the mechanism is not forbidden by crystal symmetry in these devices and the question of disentangling the two is legitimate. Four criteria separate them, and all are experimentally accessible:
\begin{enumerate}
\item \emph{Behaviour at $B\to0$.} The BCD response exists at zero field and is a property of the band structure. The effect discussed here is proportional to $B_z$ through Eq.~\eqref{eq:class}, requires Landau quantisation and incompressible strips, and vanishes identically at $B=0$.
\item \emph{Parity in $B$.} Being $\Top$-even, the BCD contribution need not be odd under field reversal; the confinement-driven term must be, Eq.~\eqref{eq:Bodd}.
\item \emph{Response to the gates.} The BCD is fixed by the lattice and is unaffected by interchanging $V_{\mathrm L}$ and $V_{\mathrm R}$; the confinement term changes sign, Eq.~\eqref{eq:gex}.
\item \emph{Order of magnitude.} The BCD in a GaAs 2DEG is governed by the Berry curvature of a nearly parabolic conduction band with weak spin--orbit coupling and is correspondingly tiny, whereas the effect computed here is of order unity in the plateau-edge region (Table~\ref{tab:resp}).
\end{enumerate}
The two are thus not competing explanations of the same data so much as responses living in different regions of parameter space, and criteria 1--3 are the same measurements already proposed in Sec.~\ref{sec:artefacts}.

There is a further reason why the two do not compete in the regime of interest. The BCD response is a Fermi-surface property of Bloch states treated in the semiclassical wave-packet approximation, and presupposes a Fermi surface over which the dipole may be integrated. At 5--7\,T and $\nu\simeq2$--3 the spectrum is fully Landau quantised, the Fermi surface has been replaced by discrete broadened levels, and the expansion that defines the dipole does not apply; the present effect, by contrast, requires precisely that quantisation, since it is carried by the incompressible strips. The two mechanisms therefore occupy disjoint regions of field rather than competing at a common operating point, and criterion 1 tests exactly this.

\subsection{Hydrodynamic and ballistic nonreciprocity}\label{sec:hydro}

Two further mechanisms of nonreciprocal transport have been discussed for high-mobility GaAs, and neither is excluded by the mobility of the devices of Refs.~\cite{Siddiki2010NJP,Horas2008}. In the hydrodynamic regime the viscosity of the electron liquid may depend linearly on the flow velocity once time reversal and Galilean invariance are both absent, which produces a nonreciprocity substantially enhanced over its disorder-dominated counterpart~\cite{Kirkinis2025}. In the ballistic regime an asymmetric channel geometry rectifies by itself, as in the ballistic rectifier of Ref.~\cite{Song1998}, which is a GaAs device whose inversion breaking is, as here, lithographic rather than crystalline.

The hydrodynamic mechanism is excluded at our operating point by a length-scale estimate. Viscous flow requires the electron--electron scattering length to be the shortest length in the problem, $l_{ee}\ll W$. At $kT=5\times10^{-3}\EFc$ ($T\simeq0.75$\,K) and $\bar n_{\mathrm{el}}\simeq3\times10^{11}$\,cm$^{-2}$, with $l_{ee}\sim v_{\mathrm F}\hbar E_{\mathrm F}/(kT)^2\ln(E_{\mathrm F}/kT)$, one finds $l_{ee}\sim10^2\,\mu$m against a bar width of $5\,\mu$m: the inequality is violated by more than an order of magnitude. Consistently, the viscous window reported for GaAs channels and point contacts of comparable width lies near 5--50\,K~\cite{Gusev2018}, two orders of magnitude above the temperature at which the plateau structure that carries the present effect survives at all.

Both mechanisms are separated from the present one by the field window in any case, and this is the more robust discriminator because it does not rest on a scattering-length estimate. Neither viscous nor ballistic rectification requires Landau quantisation; both operate on the compressible carriers directly, and neither switches off when the bulk becomes incompressible. The computed response here vanishes identically at $\nu_0=2.0$ and 2.2 (Fig.~\ref{fig:sweep}), reaches its maximum at the low-field plateau edge, and falls by more than an order of magnitude beyond the plateau. A rectification of comparable sign and magnitude surviving at $\nu_0=2.0$, or persisting to $B=0$, is not the effect described here. The temperature dependence separates them a second time and in the opposite direction: the present effect tracks the plateau edge and hence $kT/\hbar\omega_c$ (Fig.~\ref{fig:temp}), diminishing as the temperature rises, whereas a viscous contribution grows with the electron--electron scattering rate over the same range.

\subsection{Where the quasi-local description applies}\label{sec:validity}

A local relation between $\bm j$ and $\nabla\mu^*$ presupposes averaging over regions small compared with the scale on which the density varies yet containing many electrons. Two lengths bound the window from below: the strip must satisfy $W>\lambda_{\mathrm F}$, since for $\ell_B<W<\lambda_{\mathrm F}$ it becomes evanescent and current leaks into the adjacent compressible region~\cite{Sailer2010}, and for $W<\ell_B$ the wave functions bridge it. This is why the conductivity is averaged over $\lambda\sim\lambda_{\mathrm F}/2$~\cite{Siddiki2004PRB}; the averaging is the minimal repair of locality, and Sec.~\ref{sec:conv} shows that the admissible window $\ell_B\lesssim\lambda\lesssim\lambda_{\mathrm F}$ is set by the sample rather than chosen, and that varying $\lambda$ within it relocates the effect in field rather than removing it (Table~\ref{tab:lambda}). From above, the sample must be long enough for carriers to equilibrate between the current contacts, the dissipation associated with $\RH$ being localised in hot spots at the contacts.

Our calculations sit inside this window: $W\simeq56$--106\,nm against $\lambda_{\mathrm F}\simeq46$\,nm and $\ell_B\simeq10$--11\,nm over $5.4\le B\le6.6$\,T, in a $5\,\mu$m bar. They would not apply to a quantum point contact, to a bar a few $\lambda_{\mathrm F}$ wide, or at temperatures where phase coherence spans the sample --- precisely the regimes for which the LB description, the persistent-current analysis~\cite{SirtLudwig2026} and the quantum-electrostatic treatment~\cite{Armagnat2020} are the appropriate tools.

\subsection{Current-induced heating}\label{sec:heating}

The temperature is an input to the calculation, held at $kT=5\times10^{-3}\EFc$ ($T\simeq0.75$\,K here), and the Joule heat the imposed current deposits is not fed back into it. Since Fig.~\ref{fig:temp} shows the plateau edge to be sensitive to temperature, this omission needs quantifying rather than merely listing.

The dissipation is itself polarity dependent, and strongly so. At the operating point of Table~\ref{tab:resp} the longitudinal dissipation $I^2\RL$ is 12.4\,pW at $+I$ and 1.2\,fW at $-I$, a ratio of $10^4$, simply because $\RL$ differs by that factor; the contact dissipation $I^2\RH\simeq516$\,pW is the same at both polarities and is localised in the hot spots at the current contacts rather than in the strips. Two consequences follow. First, heating cannot manufacture the effect or its sign: the larger heat load falls on the polarity that is already the less well quantized, so a self-consistent treatment would reinforce the asymmetry rather than create it, and no reversal of sign can come from this quarter. Second, it would reduce the computed magnitude at a given field and displace the window. Reading Fig.~\ref{fig:temp} backwards gives the scale: raising $kT$ from $5\times10^{-3}$ to $2\times10^{-2}\EFc$, i.e.\ from 0.75 to 3.0\,K, moves the low-field plateau edge from $\nu_0=2.30$ to 2.12. An electron temperature elevated at $+I$ would therefore narrow the window from its low-field side at that polarity, and the $\Delta\RH$ we quote should be read as an upper bound in this respect.

The symmetry statements are untouched, and it is worth seeing why. Under $\Pop$ the configuration $\{B,\Vconf(x),+I\}$ maps to $\{B,\Vconf(-x),-I\}$, and the Joule source $\bm j\cdot\bm E$ is a functional of the same self-consistent solution, so it maps covariantly along with everything else: a calculation in which $T(x)$ is determined self-consistently still satisfies Eqs.~\eqref{eq:cov} and~\eqref{eq:comp}. What heating adds is one more condition on the experiment rather than on the model. The thermal boundary conditions must mirror as well as the electrostatic ones, so the two edges must be equally anchored to the bath; a device whose two gates differ in thermal contact would break Eq.~\eqref{eq:gex} through the temperature field even with the confinement exactly exchanged. This is checkable in the same way as the other residuals, by repeating the protocol at two bath temperatures: a genuine confinement-driven rectification scales with the strip geometry, whereas a thermal-asymmetry artefact tracks the bath.

A proper treatment requires an electron-temperature model coupled to a phonon bath, with the electron--phonon coupling and the device thermal conductance as further inputs. We have not attempted it, and we note that it would be a different calculation from the one reported here, not a refinement of it.

\subsection{Disorder and the plateau asymmetry}\label{sec:disorder}

A second, independent asymmetry is documented in these devices: at high mobility the plateaus at even filling factors are displaced relative to their intersection with the classical Hall line, an asymmetry that is reduced as the mobility is lowered~\cite{Horas2008,Siddiki2009EPLb}. The observable deserves stating carefully, because the classical Hall line is not an auxiliary construction drawn through the data but the high-temperature limit of $\RH$ itself, $\RH\to B/(\nel e)$, measured on the same sample. Asking where it cuts the plateau is therefore the same as asking how the plateau contracts as the temperature is raised, and it is in that form that Ref.~\cite{Horas2008} settles it: in a gate-defined $10\,\mu$m bar at $\mu=2.96\times10^6$\,cm$^2$/Vs the contraction is one-sided, $\RH$ departing from the plateau on its low-field side while the high-field end stays pinned, so that the highest-temperature trace crosses the plateau close to its high-field edge; in an identically patterned bar at $\mu\lesssim10^6$\,cm$^2$/Vs the contraction is two-sided and the crossing returns to the plateau centre, as the textbook picture assumes. (It is a different observable from the Hall resistance overshoot of Ref.~\cite{Sailer2010}, in which $\RH$ exceeds the plateau value at its low-field end.) The screening theory attributes it to the electrostatics of the incompressible region, which occupies the bulk at the high-field end of the plateau and survives as two narrowing edge strips towards lower field. The asymmetry then follows from the order in which the strips are lost. Raising the temperature destroys an incompressible region once $kT$ becomes comparable to the gap across it, and the narrow strips go first~\cite{Horas2008,Siddiki2004PRB}; since the strips are at their narrowest on the low-field side of the plateau, where the bulk region has already broken up and only two thin edge strips remain, that is the side on which quantization is lost first. The high-field end, where the incompressible region is the bulk, is the most robust. The one-sided contraction and the position of the crossing are thus two statements about the same underlying object, and neither is a property of the Landau level alone. This is precisely the field dependence of the strip geometry that Fig.~\ref{fig:sweep}(a),(b) displays directly: the pinned region occupies the whole bar at $\nu_0=2.0$ and has narrowed to 94 and 44\,nm by $\nu_0=2.6$.

\begin{figure*}[t]
\centering
\includegraphics[width=\textwidth]{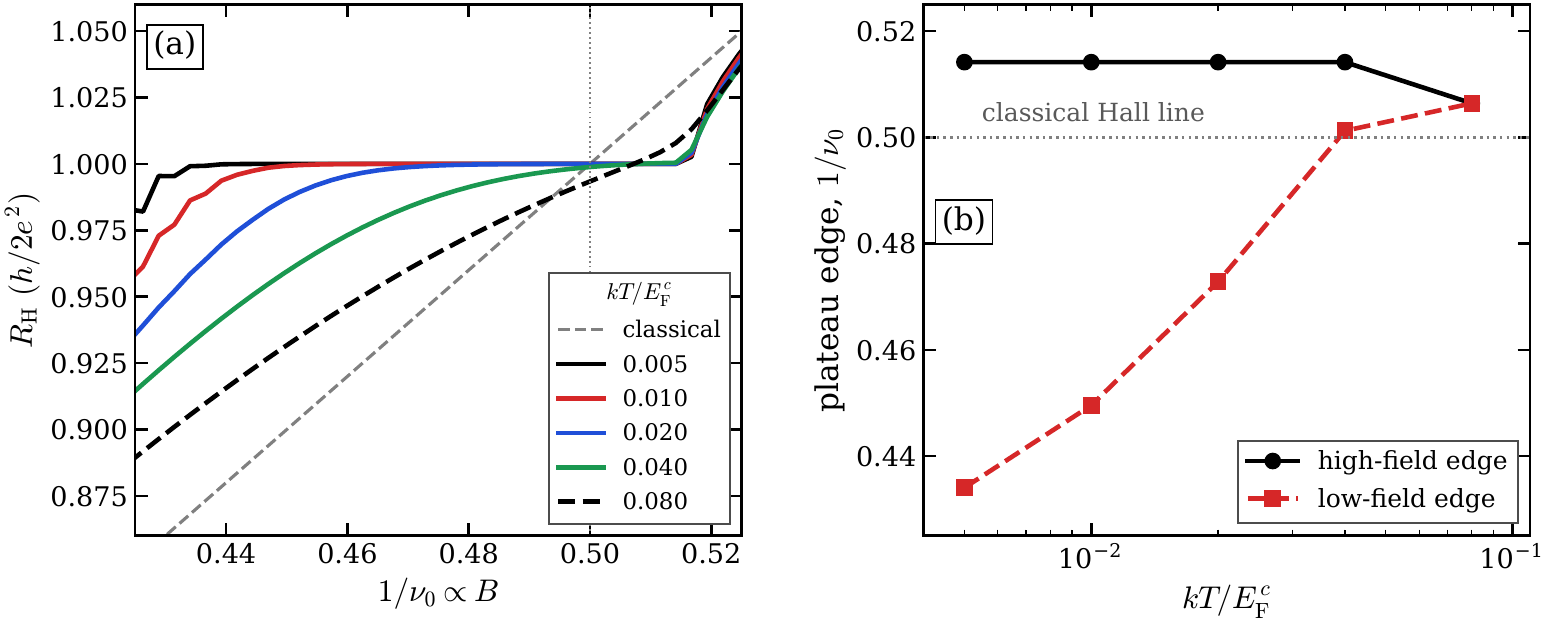}
\caption{Temperature dependence of the $\nu_0=2$ plateau in a symmetric bar (depletion $150|150$\,nm), in linear response ($e\VH=0.01\EFc$, $N=300$, $\lambda=40$\,nm), each curve a field sweep at fixed temperature. (a) $\RH$ against $1/\nu_0\propto B$ at $kT/\EFc=0.005$, 0.010, 0.020, 0.040 (solid black, red, blue, green) and 0.080 (dashed black); the grey dashed line is the classical Hall resistance $h/\nu_0e^2$, which meets the quantized value at $1/\nu_0=0.5$. Quantization is lost from the low-field side as the temperature rises, the high-field end remaining pinned. (b) The two plateau edges, located by $|\RH-1|<10^{-3}$, against temperature: the high-field edge is stationary at $1/\nu_0=0.5141$ over a factor of eight in $kT$, while the low-field edge advances from 0.4341 to 0.5012 and the two merge above $kT/\EFc\simeq0.04$. This is the one-sided contraction reported for high-mobility gate-defined bars in Ref.~\cite{Horas2008}; the mobility dependence reported there is not accessible with the uniform SCBA broadening used here.}
\label{fig:temp}
\end{figure*}

Our calculation gives this in an extreme form. Locating the plateau edges by $|\RH-1|<10^{-3}$ in the symmetric-bar sweep of Fig.~\ref{fig:temp}, the computed $\nu_0=2$ plateau at the lowest temperature runs from $1/\nu_0=0.4341$ to 0.5141, so that the classical line cuts it at $\Delta_-=0.066$ from its low-field edge and only $\Delta_+=0.014$ from its high-field edge: almost the whole plateau lies on the low-field side of the crossing. We quote the asymmetry from this sweep throughout, since it is the one whose edge criterion is stated and whose temperature dependence is resolved; Fig.~\ref{fig:asym}(a) shows the same one-sidedness at a different imposed current and is not used to fix $\Delta_\pm$. Read against Ref.~\cite{Horas2008} this is the high-mobility limit pushed past where any real sample sits, rather than a result of the wrong sign: the trend with decreasing mobility is towards symmetry, and our SCBA disorder supplies too little of whatever drives that trend. That extremeness is accordingly a warning rather than a confirmation. Reference~\cite{Siddiki2004PRB} already noted that the high-field edge of a computed plateau falls where an incompressible strip first appears at the sample centre, whereas experimentally the corresponding $\nu_0$ lies near the plateau centre, and attributed the discrepancy to the neglect of long-range potential fluctuations. We were unable to restore the experimental asymmetry with the modulation $\Vconf\to\Vconf+V_{\mathrm{mod}}\cos(m_p\pi x/d)$ of Ref.~\cite{Siddiki2009EPLb} up to $V_{\mathrm{mod}}=0.5\EFc$ and $m_p=3$--5. We think this is a defect of the model potential rather than of the mechanism, and it is worth setting out what a realistic treatment would change.

Two features of the SCBA disorder used here are responsible. It enters only as a spatially uniform level broadening $\Gamma$, and it is short ranged, the Gaussian scatterers having range $R=10$\,nm. Such a term broadens the local density of states identically at every $x$; it shifts both plateau edges together and cannot create incompressible regions where the clean profile has none. The potential of the remote ionized donors is the opposite in both respects: it is a random field with a finite correlation length, set by the spacer thickness $d_s$ separating the doping layer from the two-dimensional system and therefore of order 50--100\,nm, and with an rms amplitude fixed by the areal donor density. Its effect is not to broaden a level but to corrugate the local filling factor, $\nu(x,y)=\nu_0+\delta\nu(x,y)$ with $\delta\nu$ correlated over $d_s$.

The consequence for the plateau asymmetry is a percolation problem rather than a broadening problem. Where the clean calculation has $\nu$ passing smoothly through the integer at two points, the corrugated system has closed contours: incompressible puddles that appear wherever the local filling reaches $N$, including at average fillings $\nu_0$ slightly below $N$, i.e.\ on the high-field side, where the clean profile is entirely compressible. A quantized plateau then survives as long as such regions percolate across the bar or, in a narrow bar, as long as they bridge it. This is precisely the side on which our calculation produces nothing ($\Delta_+\lesssim0.003$), which is why we regard the one-sidedness of Fig.~\ref{fig:asym}(a) as a signature of the missing physics rather than of the symmetry. Two quantitative consequences would follow and are testable against Ref.~\cite{Siddiki2009EPLb}: the high-field edge should advance as the corrugation amplitude grows, making the plateau progressively more symmetric as mobility falls, which is the observed trend; and the asymmetry should be a function of $d_s/W$, since puddles percolate more easily when their correlation length is small compared with the bar width.

The first of these is the statement of Ref.~\cite{Horas2008} that at low mobility disorder widens the incompressible strips with respect to the magnetic field and so extends the plateau towards higher $B$; the percolation picture is that sentence given a mechanism, the widening being the appearance of incompressible puddles at average fillings for which the clean profile is entirely compressible. Implementing this requires generating Gaussian random fields with the appropriate $d_s$ and variance and averaging the plateau edges over realisations, which is a different and considerably heavier computation than the one reported here. Range-dependent disorder has been examined within the screening theory for the plateau \emph{widths}~\cite{Siddiki2007IJMPB}, but not for the asymmetry, which needs both edges resolved separately. One check is cheaper than that and is available within the present model. Because the crossing point is fixed by the order in which the incompressible regions are destroyed, the calculation should reproduce the one-sided contraction of Ref.~\cite{Horas2008} whether or not the absolute position of the plateau is right: the low-field edge should retreat with temperature while the high-field edge stays put. We have carried this out, and Fig.~\ref{fig:temp} shows the result. Over a factor of eight in temperature, $kT/\EFc=0.005$ to 0.040, the high-field edge does not move at all, sitting at $1/\nu_0=0.5141$ to within the field resolution of the sweep, while the low-field edge advances from 0.4341 to 0.5012 and the plateau width falls from 0.080 to 0.013. The contraction is entirely one-sided, and it is one-sided in the direction reported in Ref.~\cite{Horas2008}. Read against the classical line, the low-temperature plateau is cut at $\Delta_-=0.066$ from its low-field edge and $\Delta_+=0.014$ from its high-field edge --- close to the high-field end, as observed in the high-mobility sample --- and by $kT/\EFc=0.040$ the plateau has retreated past the classical line altogether.

This is the outcome that confines the defect. The mechanism --- narrow strips dying first, so that quantization is lost from the low-field side --- is reproduced without adjustment, so the strip electrostatics is not where the model fails. What it does not reproduce is the \emph{mobility} dependence, and that is a statement about the disorder potential alone: a uniform SCBA broadening has no mobility-dependent effect on which side the plateau contracts from, because it shifts both edges together. The percolation account above is therefore not an appeal to unspecified missing physics but a prediction about the one ingredient the temperature test leaves untested. We accordingly regard the quantitative account of the mobility-dependent plateau asymmetry as open, while noting that its temperature-dependent counterpart is not, and we do not use Fig.~\ref{fig:asym} as evidence for any symmetry statement.

\subsection{Onsager--Casimir reciprocity}

Onsager--Casimir reciprocity, $G_{ij}(B)=G_{ji}(-B)$, is a linear-response statement resting on microscopic reversibility. It does not force the finite-bias resistance to be even in $I$ at fixed $B$; the rectification discussed here lies outside its premises, since the current reshapes the self-consistent electrostatic state through $\mu^*(x)$ in Eq.~\eqref{eq:n}. Reciprocity is not violated. Equations~\eqref{eq:cov}--\eqref{eq:Bodd} do not rely on it either: they follow from Eq.~\eqref{eq:Keven} and the covariance of Eqs.~\eqref{eq:V}--\eqref{eq:R} under a relabelling of coordinates. This is the practical reason for distinguishing $\Pop$ from $\Mop\Top$ in Eq.~\eqref{eq:MT}.

\subsection{Relation to gauge-invariant nonlinear transport theory}

The requirement that a nonlinear response be computed with the current-induced internal potential determined self-consistently is not peculiar to the screening theory. In the mesoscopic scattering formalism the same requirement appears as the condition of gauge invariance for the nonlinear conductance coefficients: a uniform shift of all reservoir potentials together with the internal potential must produce no current, which forces the internal potential to be determined self-consistently from the injected charge rather than held at its equilibrium value~\cite{Christen1996}. S\'anchez and B\"uttiker showed that it is this screening response which makes the nonlinear current--voltage characteristic asymmetric in the magnetic field, and illustrated it for a quantum Hall bar with an antidot~\cite{Sanchez2004}; the interaction origin of the asymmetry was established in parallel in Ref.~\cite{Spivak2004}.

The present calculation is the semiclassical, self-consistent-electrostatic counterpart of that statement, in a regime where the internal potential is not a single parameter but the whole profile $\mu^*(x)$ of Eqs.~\eqref{eq:V}--\eqref{eq:n}. The polarity information is carried by the current-induced redistribution of that profile: a calculation that froze the electrostatics at its equilibrium form would return $\Delta R_\alpha\equiv0$ whatever the confinement, which is another way of seeing why the ideal edge-channel response of Sec.~\ref{sec:LB} gives zero.

Two differences of substance should be recorded. First, the relations derived here are exact rather than leading order in the bias: Eq.~\eqref{eq:cov} follows from the covariance of Eqs.~\eqref{eq:V}--\eqref{eq:R} under a relabelling of coordinates, not from a weak-nonlinearity expansion, and it therefore survives into the strongly nonlinear plateau-edge regime where $\Delta R_\alpha$ is of order unity. Second, the symmetry exploited here is a spatial one imposed by the gates, so that Eq.~\eqref{eq:comp} fixes the field-reversal behaviour of the response with a definite sign, whereas in a mesoscopic conductor the corresponding asymmetry is a sample-specific fluctuating quantity.

\section{What the argument does not cover}\label{sec:notcover}

Spin is omitted from the numerics, and every quantitative statement in this paper --- Tables~\ref{tab:resp}--\ref{tab:sens}, Figs.~\ref{fig:profile}--\ref{fig:RL}, the strip widths, the resistances and the nominal coefficient $\gamma$ --- refers to even filling factors of a spin-degenerate calculation with $g^*=0$. The symmetry results are unaffected: as noted in Sec.~\ref{sec:model}, $H_{\mathrm Z}$ is covariant under $\Top$, $\Mop$ and $\Pop$ because spin and $\bm B$ are both axial, so Eqs.~\eqref{eq:cov}--\eqref{eq:Bodd} and the classification~\eqref{eq:class} hold with spin included.

What changes at odd $\nu$ is the energy scale that stabilises the strip, and hence the magnitude and field window of the rectification. The gap bounding an incompressible region is $\hbar\omega_c-g^*\mu_{\mathrm B}B$ at even $\nu$ but the Zeeman gap $g^*\mu_{\mathrm B}B$ at odd $\nu$, which in GaAs is smaller by more than an order of magnitude in its bare form. Exchange enhancement compensates part of this: the effective $g^*$ is strongly increased inside an incompressible strip, where the spin polarisation is complete, and the self-consistent treatments of Refs.~\cite{Siddiki2008PhysE,Bilgec2010} produce odd-integer strips of finite width on that basis. Three consequences follow for the present effect, none of which we have computed. The strips at odd $\nu$ are narrower for the same confinement, so the current-induced widening of Eq.~\eqref{eq:W} --- which scales as $\sqrt{\Delta E_{\mathrm{eff}}}$ --- operates on a smaller gap and reaches the collapse condition $\Delta E_{\mathrm{eff}}\to0$ at a smaller $e\VH$, so the rectification should set in at lower current. The field window in which it is visible should correspondingly narrow, since the plateau itself is narrower. And because the exchange enhancement is itself a function of the local filling, the widening should be less symmetric between the two polarities than in the spin-degenerate case, the self-consistent gap responding to the current as well as the width. None of this changes the sign: that is fixed by which edge carries the wider strip and by the sense of the current-induced tilt of $\mu^*$, both properties of the confinement, whereas exchange enhancement alters the gap $\Delta E_{\mathrm{eff}}$ entering Eq.~\eqref{eq:W}, and hence the current at which a strip collapses, but not which of the two strips is the wider. We therefore expect the sign and the transformation properties of Sec.~\ref{sec:ops} to carry over to odd $\nu$ unchanged, with the operating current and the width of the field window as the quantities that shift. We note that the measured $\nu=1$ plateau of Ref.~\cite{Siddiki2009EPLb} is more symmetric than the even ones, which is consistent with a more robust, exchange-stabilised bulk strip at that filling; whether the same holds for the polarity asymmetry is an open question that a spin-resolved calculation within the same framework could answer. Contacts, probe geometry and finite length lie outside the derivation, and $\Pop$ relabels the voltage probes, so the relations apply to contact configurations that map onto each other under $(x,y)\to(-x,-y)$. Local heating is not treated self-consistently~\cite{Sailer2010,Siddiki2010NJP}; scanning-force measurements approaching breakdown~\cite{Panos2014} show how much structure that omission discards. Consequently the calculation identifies which polarity retains the wider current-carrying strip, and hence which is more robust, but does not predict the absolute breakdown current.

\section{Conclusion}

The quantized Hall effect is not indifferent to which way the current flows, nor to how the electrons are confined. Its plateau value is; its plateaus are not. A bar whose two edges differ is a rectifier at the plateau edge, and the direction it favours is fixed by the gates, not by the sample's history or by chance. That is a claim about where the universality of the effect ends, and it is a claim that can be checked on existing devices.

For a two-dimensional electron system in a perpendicular field, the relevant nontrivial discrete operation that reverses the longitudinal current while leaving $B_z$ invariant is the in-plane $C_2$ rotation $\Pop$, which coincides in its action on the configuration with $\Mop\Top$. (Proper rotations and translations also leave a perpendicular field invariant; what singles out $\Pop$ is that it reverses $I$ and exchanges the two edges once the bar geometry is specified.) Within the self-consistent quasi-local screening model $\Pop$ yields the covariance relation~\eqref{eq:cov}, from which two experimental statements follow: the current-polarity rectification of a hard--soft bar is odd under exchange of the gate voltages and vanishes for a fully mirror-symmetric configuration, and its antisymmetric component reverses when the magnetic field is reversed. Both are consequences of the model, not requirements imposed on the data: each is exact within the quasi-local screening description under mirror-related boundary conditions, and each carries the residuals set out in Secs.~\ref{sec:artefacts} and~\ref{sec:heating} --- contact placement along the bar, thermal anchoring of the two edges, and the device drift that a magnet sweep can introduce. The first has been observed~\cite{Siddiki2010NJP}; the second has not been tested. Taken together they define the measurement we would most like to see performed. Eight traces --- both current polarities, both field directions, both gate configurations --- suffice to isolate the component odd under all of them, and it is that double-parity component, not the size of the rectification, that the symmetry argument predicts. The rectification itself has been observed and motivates the paper; the double-parity experiment is what the paper is for, and it is the decisive test of the symmetry statement.

A relation of this kind constrains without discriminating, however, and we have been explicit about it in Secs.~\ref{sec:artefacts} and~\ref{sec:LB}: the ideal edge-channel response satisfies Eq.~\eqref{eq:cov} trivially, by not depending on the confinement at all. The eight-trace protocol therefore establishes that the effect is a confinement-driven orbital nonreciprocity and not a thermovoltage, a contact drift or a Berry-curvature-dipole term; it does not by itself decide between the two descriptions of where the current flows. That is the separate job of the strip-collapse measurement of Sec.~\ref{sec:collapse}, in which one edge is made progressively harder and the co-directed and chiral pictures predict saturation against growth. The two experiments are complementary and we would run both: the first settles what the effect is, the second what carries it. The symmetry of the effect is that of a vector-product nonreciprocity, $\delta R\propto B_z(\bm p\times\bm I)_z$, with the inversion breaking supplied by the gates rather than by a crystal --- a nonreciprocal response that is tunable, reversible and not of relativistic origin.

\backmatter

\bmhead{Acknowledgements}

I thank R.~R.~Gerhardts, with whom the self-consistent screening description of the integer quantized Hall effect used throughout this work was developed, and S.~Ludwig, J.~Horas, D.~Kupidura and W.~Wegscheider for the experiments on gate-defined asymmetric Hall bars that motivated the present symmetry analysis. Discussions with S.~Sirt and S.~Ludwig on the decomposition of the current density into imposed and persistent components, on which we deliberately take no position here, sharpened the statement of Sec.~\ref{sec:model}. I am grateful to K.~von Klitzing and J.~Weis for many discussions on the current and potential distribution in the quantum Hall regime.

\section*{Declarations}

\begin{itemize}
\item \textbf{Funding.} This research did not receive any specific grant from funding agencies in the public, commercial, or not-for-profit sectors.
\item \textbf{Competing interests.} The author has no relevant financial or non-financial interests to disclose.
\item \textbf{Data availability.} The data underlying every figure and table, together with the input files used to generate them, are available from the corresponding author on reasonable request and will be deposited in a public repository upon acceptance.
\item \textbf{Code availability.} The Python implementation of the \scblin\ code used for all calculations, and the plotting scripts, are available from the corresponding author on reasonable request and will be deposited in a public repository upon acceptance.
\item \textbf{Author contribution.} A.S. is the sole author: he conceived the study, developed the theory, designed and analysed the calculations, and wrote the manuscript.
\item \textbf{Use of generative AI and AI-assisted technologies.} During the preparation of this work the author used the following tools. Claude (Anthropic) translated the original Fortran source of the \scblin\ code into Python, ran the calculations, and wrote the scripts that produce the figures directly from the computed data; the translation and its verification are described in Appendix~\ref{app:num}. ChatGPT (OpenAI) was used for a critical pre-submission review of the manuscript. Jenni AI was used to check the references and to review the manuscript. Grammarly was used for English-language polishing. After using these tools the author reviewed and edited the code, the results and the text, and takes full responsibility for the content of the publication.
\end{itemize}

\begin{appendices}
\section{Effective particle--hole transformation and its limitations}\label{app:C}

Charge conjugation was excluded from the main text because two distinct operations travel under that name and they are not equivalent in a gate-defined GaAs/AlGaAs device.

$\mathcal C_{\mathrm{em}}$ is the field-theoretic operation: $e\to-e$ for all charges, including those generating the field and biasing the gates, so $\bm E\to-\bm E$, $\bm B\to-\bm B$, $I\to-I$. It maps the heterostructure, its donors, gates and reservoirs onto a different physical system, and it reverses $B$.

$\mathcal C_{\mathrm{ph}}$ is an effective particle--hole transformation within a Landau-level description, $\nu(x)\to\nu_c-\nu(x)$ at fixed $B$ and fixed gates. It is not an exact symmetry of Eq.~\eqref{eq:H}: its very definition depends on the Landau-level projection, on the reference filling $\nu_c$, and on the treatment of disorder, confinement and Landau-level mixing. Statements about $\mathcal C_{\mathrm{em}}$ do not transfer to $\mathcal C_{\mathrm{ph}}$, or conversely.

Within the SCBA the local conductivities are particle--hole symmetric, since the spectral functions $A_n(E)=(2/\pi\Gamma_n)\sqrt{1-[(E-E_n)/\Gamma_n]^2}$ are symmetric about $E_n$. The Hartree term is not. Using Eq.~\eqref{eq:V},
\begin{equation}
\begin{split}V_{\mathrm H}[\nu_c-\nu]={}&-V_{\mathrm H}[\nu]\\&+\frac{2e^2}{\bar\kappa}\int dx'\,K(x,x')\bigl[n_c-2n_0\bigr],\end{split}
\label{eq:Cph}
\end{equation}
the comparison being at fixed electrochemical potential; since $\nu\to\nu_c-\nu$ does not conserve particle number, at fixed $N$ an additional shift of $\mu^*$ appears, which changes the constant but not the fact that it is non-zero. The Hartree term is thus particle--hole symmetric only for the fine-tuned background $n_0=n_c/2$, and even there $\mathcal C_{\mathrm{ph}}$ would be broken by the confinement, since a potential that confines electrons expels holes.

Combinations follow directly. $\mathcal C_{\mathrm{em}}\Mop$, $\mathcal C_{\mathrm{em}}\Top$, $\mathcal C_{\mathrm{em}}\Pop$ and $\mathcal C_{\mathrm{em}}\Mop\Top$ all fail for a fixed device, the last two because they reverse $B$. Nothing about the relativistic CPT theorem follows from this, and we make no such claim: that theorem presupposes a Lorentz-invariant, local, unitary field theory of a closed system, whereas a Hall bar is non-relativistic, driven, dissipative and coupled to reservoirs, with $B$ and $\Vconf$ external classical parameters. (In 2+1 dimensions the corresponding statement is in any case a CRT rather than a CPT one, since $(x,y)\to(-x,-y)$ has $\det=+1$~\cite{Streater1964,Deser1982}.) The statements here concern the invariance group of a fixed experimental configuration and nothing more.

Finally, the relation between $\mathcal C_{\mathrm{ph}}$ and the measured plateau asymmetry is weaker than that between $\Pop$ and the rectification. Equation~\eqref{eq:DR2} makes $\Delta R_\alpha$ vanish for a fully mirror-symmetric configuration; nothing so tight holds for $\mathcal C_{\mathrm{ph}}$, since a symmetric plateau would not imply unbroken $\mathcal C_{\mathrm{ph}}$ --- disorder, temperature and the finite extent of the wave functions all shift the plateau edges independently. The plateau asymmetry is a diagnostic, not an order parameter, and Sec.~\ref{sec:disorder} explains why we do not press it further.

\section{Numerical procedure}\label{app:num}

All results were produced with the \scblin\ program~\cite{SCBLIN}, implementing the scheme of Ref.~\cite{Siddiki2004PRB}. The original Fortran implementation was translated into Python with the assistance of Claude (Anthropic), which also executed the calculations reported here and generated the plotting scripts. The translation was verified against the Fortran original by reproducing the Poisson kernel to machine precision against adaptive quadrature of the original integrands and by reproducing the published density and resistance traces of Ref.~\cite{Siddiki2004PRB}; every figure is plotted by script directly from the computed data, without manual adjustment. The sequence is: (i) the linear integral equation for the density profile is solved at $T=0$, $B=0$ on a mesh of $N$ intervals spanning $|x|\le d$, the Poisson kernel~\eqref{eq:K} being integrated panel by panel against quadratic Lagrange stencils, with the end panels regularised by $t=\mp(1-\tfrac h2u^2)$ to remove the $\sqrt{1-t^2}$ branch point; (ii) the result is continued to finite temperature at $B=0$ by Newton--Raphson; (iii) the field is switched on and the temperature lowered in steps to the target value, the density being recomputed from Eq.~\eqref{eq:n} with SCBA spectral functions at each step; (iv) the current is imposed and the electrochemical potential from Eqs.~\eqref{eq:ohm}--\eqref{eq:R} fed back into Eq.~\eqref{eq:n}, the loop repeating until $\max_x|V^{(k+1)}-V^{(k)}|<10^{-8}E_0$. Conductivities are the SCBA ones for Gaussian scatterers of range $R$; both they and the filling factor are averaged over $\lambda$ before entering Eq.~\eqref{eq:R}.

Figure~\ref{fig:profile} uses $N=800$ (6.25\,nm) and Fig.~\ref{fig:asym} $N=300$. The lateral asymmetry enters through the two depletion lengths and the current polarity through the sign of the target Hall voltage. The mirror test~\eqref{eq:mirrortest} exchanged the two depletion lengths and the sign of the Hall voltage simultaneously and compared converged profiles point by point. Table~\ref{tab:sens} was generated by varying one input at a time about the reference row.

\end{appendices}


\begin{thebibliography}{99}

\bibitem{Buttiker1988} M.~B\"uttiker, Phys. Rev. B 38 (1988) 9375. \href{https://doi.org/10.1103/physrevb.38.9375}{doi:10.1103/physrevb.38.9375}

\bibitem{Siddiki2004PRB} A.~Siddiki, R.~R.~Gerhardts, Incompressible strips in dissipative Hall bars as origin of quantized Hall plateaus, Phys. Rev. B 70 (2004) 195335. \href{https://doi.org/10.1103/physrevb.70.195335}{doi:10.1103/physrevb.70.195335}

\bibitem{Siddiki2004IJMPB} A.~Siddiki, R.~R.~Gerhardts, The interrelation between incompressible strips and quantized Hall plateaus, Int. J. Mod. Phys. B 18 (2004) 3541. \href{https://doi.org/10.1142/s0217979204026974}{doi:10.1142/s0217979204026974}

\bibitem{Guven2003} K.~G\"uven, R.~R.~Gerhardts, Phys. Rev. B 67 (2003) 115327. \href{https://doi.org/10.1103/physrevb.67.115327}{doi:10.1103/physrevb.67.115327}

\bibitem{Gerhardts2008} R.~R.~Gerhardts, The effect of screening on current distribution and conductance quantisation in narrow quantum Hall systems, Phys. Status Solidi B 245 (2008) 378. \href{https://doi.org/10.1002/pssb.200743344}{doi:10.1002/pssb.200743344}

\bibitem{Ahlswede2001} E.~Ahlswede, P.~Weitz, J.~Weis, K.~von Klitzing, K.~Eberl, Hall potential profiles in the quantum Hall regime measured by a scanning force microscope, Physica B 298 (2001) 562. \href{https://doi.org/10.1016/s0921-4526(01)00383-0}{doi:10.1016/s0921-4526(01)00383-0}

\bibitem{Ilani2004} S.~Ilani, J.~Martin, E.~Teitelbaum, J.~H.~Smet, D.~Mahalu, V.~Umansky, A.~Yacoby, The microscopic nature of localization in the quantum Hall effect, Nature 427 (2004) 328. \href{https://doi.org/10.1038/nature02230}{doi:10.1038/nature02230}

\bibitem{Sirt2025APL} S.~Sirt, V.~Y.~Umansky, A.~Siddiki, S.~Ludwig, Transition from edge- to bulk-currents in the quantum Hall regime, Appl. Phys. Lett. 126 (2025) 243101. \href{https://doi.org/10.1063/5.0275599}{doi:10.1063/5.0275599}

\bibitem{Siddiki2009EPL} A.~Siddiki, Current-direction induced rectification effect on (integer) quantized Hall plateaus, Europhys. Lett. 87 (2009) 17008. \href{https://doi.org/10.1209/0295-5075/87/17008}{doi:10.1209/0295-5075/87/17008}

\bibitem{Siddiki2010NJP} A.~Siddiki, J.~Horas, D.~Kupidura, W.~Wegscheider, S.~Ludwig, Asymmetric nonlinear response of the quantized Hall effect, New J. Phys. 12 (2010) 113011. \href{https://doi.org/10.1088/1367-2630/12/11/113011}{doi:10.1088/1367-2630/12/11/113011}

\bibitem{Siddiki2007IJMPB} A.~Siddiki, R.~R.~Gerhardts, Range-dependent disorder effects on the plateau-widths calculated within the screening theory of the IQHE, Int. J. Mod. Phys. B 21 (2007) 1362. \href{https://doi.org/10.1142/s0217979207042847}{doi:10.1142/s0217979207042847}

\bibitem{Siddiki2008PhysE} A.~Siddiki, The spin-split incompressible edge states within empirical Hartree approximation at intermediately large Hall samples, Physica E 40 (2008) 1124. \href{https://doi.org/10.1016/j.physe.2007.08.153}{doi:10.1016/j.physe.2007.08.153}

\bibitem{Siddiki2009EPLb} A.~Siddiki, J.~Horas, J.~Moser, W.~Wegscheider, S.~Ludwig, Interaction-mediated asymmetries of the quantized Hall effect, Europhys. Lett. 88 (2009) 17007. \href{https://doi.org/10.1209/0295-5075/88/17007}{doi:10.1209/0295-5075/88/17007}

\bibitem{Iwakiri2023} S.~Iwakiri, L.~V.~Ginzburg, M.~P.~R\"o\"osli, Y.~Meir, A.~Stern, C.~Reichl, M.~Berl, W.~Wegscheider, T.~Ihn, K.~Ensslin, Nonlinear response of a two-dimensional electron gas in the quantum Hall regime, Phys. Rev. Research 5 (2023) L032046. \href{https://doi.org/10.1103/physrevresearch.5.l032046}{doi:10.1103/physrevresearch.5.l032046}

\bibitem{SirtLudwig2026} S.~Sirt, S.~Ludwig, Current density distribution for the quantum Hall effect, New J. Phys. 28 (2026) 053503. \href{https://doi.org/10.1088/1367-2630/ae66a9}{doi:10.1088/1367-2630/ae66a9}

\bibitem{Sirt2025NJP} S.~Sirt, M.~Kamm, V.~Y.~Umansky, S.~Ludwig, Nature of current flow in the regime of the quantum Hall effect, New J. Phys. 27 (2025) 083507. \href{https://doi.org/10.1088/1367-2630/adfc05}{doi:10.1088/1367-2630/adfc05}

\bibitem{Armagnat2020} P.~Armagnat, X.~Waintal, Reconciling edge states with compressible stripes in a ballistic mesoscopic conductor, J. Phys. Mater. 3 (2020) 02LT01. \href{https://doi.org/10.1088/2515-7639/ab7582}{doi:10.1088/2515-7639/ab7582}

\bibitem{Li2024} C.~Li \emph{et al.}, Observation of giant non-reciprocal charge transport from quantum Hall states in a topological insulator, Nat. Mater. 23 (2024) 1208. \href{https://doi.org/10.1038/s41563-024-01874-4}{doi:10.1038/s41563-024-01874-4}

\bibitem{Chklovskii1992} D.~B.~Chklovskii, B.~I.~Shklovskii, L.~I.~Glazman, Phys. Rev. B 46 (1992) 4026. \href{https://doi.org/10.1103/physrevb.46.4026}{doi:10.1103/physrevb.46.4026}

\bibitem{SCBLIN} Program \scblin\ (self-consistent screening with SCBA conductivities and imposed dissipative current), the code underlying Refs.~\cite{Siddiki2004PRB,Siddiki2004IJMPB}. The calculations reported here were run with a Python translation of the original Fortran source, prepared with the assistance of Claude (Anthropic) and verified against it by reproducing the Poisson kernel to machine precision against adaptive quadrature of the original integrands and by reproducing the published density and resistance traces of Ref.~\cite{Siddiki2004PRB}. The source, the input files for every figure and table, and the version identifier will be deposited in a public archive on acceptance.

\bibitem{Horas2008} J.~Horas, A.~Siddiki, J.~Moser, W.~Wegscheider, S.~Ludwig, Investigations on unconventional aspects in the quantum Hall regime of narrow gate defined channels, Physica E 40 (2008) 1130. \href{https://doi.org/10.1016/j.physe.2007.08.160}{doi:10.1016/j.physe.2007.08.160}

\bibitem{Rikken2001} G.~L.~J.~A.~Rikken, J.~F\"olling, P.~Wyder, Electrical magnetochiral anisotropy, Phys. Rev. Lett. 87 (2001) 236602. \href{https://doi.org/10.1103/physrevlett.87.236602}{doi:10.1103/physrevlett.87.236602}

\bibitem{Rikken2005} G.~L.~J.~A.~Rikken, P.~Wyder, Phys. Rev. Lett. 94 (2005) 016601. \href{https://doi.org/10.1103/physrevlett.94.016601}{doi:10.1103/physrevlett.94.016601}

\bibitem{Ideue2017} T.~Ideue \emph{et al.}, Bulk rectification effect in a polar semiconductor, Nat. Phys. 13 (2017) 578. \href{https://doi.org/10.1038/nphys4056}{doi:10.1038/nphys4056}

\bibitem{Du2021} Z.~Z.~Du, H.-Z.~Lu, X.~C.~Xie, Nonlinear Hall effects, Nat. Rev. Phys. 3 (2021) 744. \href{https://doi.org/10.1038/s42254-021-00359-6}{doi:10.1038/s42254-021-00359-6}

\bibitem{SuarezRodriguez2025} M.~Su\'arez-Rodr\'iguez, F.~de~Juan, I.~Souza, M.~Gobbi, F.~Casanova, L.~E.~Hueso, Nonlinear transport in noncentrosymmetric systems, Nat. Mater. 24 (2025) 1005. \href{https://doi.org/10.1038/s41563-025-02261-3}{doi:10.1038/s41563-025-02261-3}

\bibitem{Kirkinis2025} E.~Kirkinis, L.~Bonds, A.~Levchenko, A.~V.~Andreev, Nonreciprocity of hydrodynamic electron transport in noncentrosymmetric conductors, Phys. Rev. Lett. 135 (2025) 106302. \href{https://doi.org/10.1103/nx9z-vkk2}{doi:10.1103/nx9z-vkk2}

\bibitem{Wakatsuki2017} R.~Wakatsuki \emph{et al.}, Nonreciprocal charge transport in noncentrosymmetric superconductors, Sci. Adv. 3 (2017) e1602390. \href{https://doi.org/10.1126/sciadv.1602390}{doi:10.1126/sciadv.1602390}

\bibitem{Sodemann2015} I.~Sodemann, L.~Fu, Quantum nonlinear Hall effect induced by Berry curvature dipole in time-reversal invariant materials, Phys. Rev. Lett. 115 (2015) 216806. \href{https://doi.org/10.1103/physrevlett.115.216806}{doi:10.1103/physrevlett.115.216806}

\bibitem{Ortix2021} C.~Ortix, Nonlinear Hall effect with time-reversal symmetry: theory and material realizations, Adv. Quantum Technol. 4 (2021) 2100056. \href{https://doi.org/10.1002/qute.202100056}{doi:10.1002/qute.202100056}

\bibitem{Song1998} A.~M.~Song, A.~Lorke, A.~Kriele, J.~P.~Kotthaus, W.~Wegscheider, M.~Bichler, Nonlinear electron transport in an asymmetric microjunction: a ballistic rectifier, Phys. Rev. Lett. 80 (1998) 3831. \href{https://doi.org/10.1103/physrevlett.80.3831}{doi:10.1103/physrevlett.80.3831}

\bibitem{Gusev2018} G.~M.~Gusev, A.~D.~Levin, E.~V.~Levinson, A.~K.~Bakarov, Viscous electron flow in mesoscopic two-dimensional electron gas, AIP Adv. 8 (2018) 025318. \href{https://doi.org/10.1063/1.5020763}{doi:10.1063/1.5020763}

\bibitem{Sailer2010} J.~Sailer, A.~Wild, V.~Lang, A.~Siddiki, D.~Bougeard, Quantum Hall resistance overshoot in two-dimensional electron gases: theory and experiment, New J. Phys. 12 (2010) 113033. \href{https://doi.org/10.1088/1367-2630/12/11/113033}{doi:10.1088/1367-2630/12/11/113033}

\bibitem{Christen1996} T.~Christen, M.~B\"uttiker, Gauge-invariant nonlinear electric transport in mesoscopic conductors, Europhys. Lett. 35 (1996) 523. \href{https://doi.org/10.1209/epl/i1996-00145-8}{doi:10.1209/epl/i1996-00145-8}

\bibitem{Sanchez2004} D.~S\'anchez, M.~B\"uttiker, Magnetic-field asymmetry of nonlinear mesoscopic transport, Phys. Rev. Lett. 93 (2004) 106802. \href{https://doi.org/10.1103/physrevlett.93.106802}{doi:10.1103/physrevlett.93.106802}

\bibitem{Spivak2004} B.~Spivak, A.~Zyuzin, Signature of the electron-electron interaction in the magnetic-field dependence of nonlinear I--V characteristics in mesoscopic systems, Phys. Rev. Lett. 93 (2004) 226801. \href{https://doi.org/10.1103/physrevlett.93.226801}{doi:10.1103/physrevlett.93.226801}

\bibitem{Bilgec2010} G.~Bilge\c{c}, H.~\"Ust\"unel Toffoli, A.~Siddiki, I.~Sokmen, The self-consistent calculation of exchange enhanced odd integer quantized Hall plateaus within Thomas--Fermi--Dirac approximation, Physica E 42 (2010) 1058. \href{https://doi.org/10.1016/j.physe.2009.11.052}{doi:10.1016/j.physe.2009.11.052}

\bibitem{Panos2014} K.~Panos, R.~R.~Gerhardts, J.~Weis, K.~von Klitzing, Current distribution and Hall potential landscape towards breakdown of the quantum Hall effect, New J. Phys. 16 (2014) 113071. \href{https://doi.org/10.1088/1367-2630/16/11/113071}{doi:10.1088/1367-2630/16/11/113071}

\bibitem{Streater1964} R.~F.~Streater, A.~S.~Wightman, PCT, Spin and Statistics, and All That, Benjamin, New York, 1964.

\bibitem{Deser1982} S.~Deser, R.~Jackiw, S.~Templeton, Three-dimensional massive gauge theories, Phys. Rev. Lett. 48 (1982) 975. \href{https://doi.org/10.1103/physrevlett.48.975}{doi:10.1103/physrevlett.48.975}

\end{thebibliography}
\end{document}